\documentclass[11pt,a4paper,english]{article}
\usepackage[titletoc, title]{appendix}
\usepackage{amsmath}
\usepackage{amssymb}
\usepackage{array}
\usepackage{babel}
\usepackage{bbding}
\usepackage{color}
\usepackage[normal]{caption}
\usepackage{subcaption}
\usepackage{epsfig}
\usepackage{graphicx}
\usepackage{psfrag}
\usepackage{proofapnd}
\usepackage[round]{natbib}
\usepackage[margin=2cm]{geometry} 
\usepackage{latexsym}
\usepackage{setspace}
\usepackage{slashbox}
\usepackage{enumitem}
\usepackage{booktabs}
\usepackage{tabularx}
\usepackage{longtable,booktabs}

\setlist[itemize]{leftmargin=*}
\setlist[description]{leftmargin=*}

\title{\LARGE \bf Copula-Based Factor Model for Credit Risk Analysis\footnote{This is a post-peer-review, pre-copyedit version of an article published in Review of Quantitative Finance and Accounting. The final authenticated version is available online at: https://doi.org/10.1007/s11156-016-0613-x}}

\author{
	\begin{tabular}[t]{ccc}
		Lu, Meng-Jou \thanks{Department of Information and Finance
			Management, Institute of Finance and Institute of Information
			Management, National Chiao-Tung University, No.1001 Daxue Rd,
			Hsinchu City, Taiwan. Ladislaus von Bortkiewicz Chair of Statistics,
			Humboldt--Universit\"{a}t zu Berlin, C.A.S.E. -- Center for
			Applied Statistics and Economics, Unter den Linden 6, 10099 Berlin, Germany. E-mail: \texttt{mangrou@gmail.com}.}
		\and\and Chen, Cathy Yi-Hsuan\thanks{Corresponding author. Department of Finance, Chung
			Hua University, 707, WuFu Rd., Hsinchu 300, Taiwan. Ladislaus von Bortkiewicz Chair of Statistics,
			Humboldt--Universit\"{a}t zu Berlin, C.A.S.E. -- Center for
			Applied Statistics and Economics, Unter den Linden 6, 10099 Berlin, Germany. E-mail:
			\texttt{cathy1107@gmail.com}. } \and\and H\"{a}rdle, Karl Wolfgang  
		\thanks{ Ladislaus von Bortkiewicz Chair of Statistics,
			Humboldt--Universit\"{a}t zu Berlin, C.A.S.E. -- Center for
			Applied Statistics and Economics, Unter den Linden 6, 10099 Berlin, Germany. Sim Kee Boon Institute for
			Financial Economics, Singapore Management University
			Administration Building, 81 Victoria Street, 188065 Singapore.
			E-mail: \texttt{haerdle@hu-berlin.de}. }
	\end{tabular}
}
\date{This version: \today}

\renewcommand{\baselinestretch}{1.2}

\begin{document}

\newtheorem{lemma}{Lemma}
\newtheorem {proposition}[lemma]{Proposition}
\newtheorem {corollary}{Corollary}
\newtheorem {theorem}{Theorem}
\newtheorem{claim}[lemma]{Claim}
\newtheorem{comment}[lemma]{Comment}
\newtheorem{example}[lemma]{Example}
\newtheorem{fact}[lemma]{Fact}
\newtheorem{defn}[lemma]{Definition}
\newtheorem{exercise}{Exercise}[section]
\newtheorem{programming}[exercise]{Programming assignment}
\newenvironment{proof}{{\flushleft\textbf{\textsl{Proof.\ \ }}}}{\hfill{\hfill\rule{2mm}{2mm}}}
\pagenumbering{arabic}
\maketitle
\begin{abstract}
\footnotesize{A standard quantitative method to access credit risk employs a factor model based on joint multivariate normal distribution properties. By extending a one-factor Gaussian copula model to make a more accurate default forecast, this paper proposes to incorporate a state-dependent recovery rate into the conditional factor loading, and model them by sharing a unique common factor. The common factor governs the default rate and recovery rate simultaneously and creates their association implicitly. In accordance with Basel III, this paper shows that the tendency of default is more governed by systematic risk rather than idiosyncratic risk during a hectic period. Among the models considered, the one with random factor loading and a state-dependent recovery rate turns out to be the most superior on the default prediction.\\
\renewcommand{\baselinestretch}{1.2}

}\pagestyle{empty}\\

\end{abstract}

\clearpage
\section{Introduction}

The global economy has repeatedly observed clusters of default
events, such as the burst of the dotcom bubble in 2001, and the
financial crisis from 2007 to 2009. The clustered default has been
attributed to systematic risk which plays a crucial role in the
default event. To discover this issue, numerous studies emphasise
the role of systematic risk by employing a factor model
({\color{blue} \citealp{andersen2004extensions}}; {\color{blue}
\citealp{pan2008default}}; {\color{blue}
\citealp{rosen2010risk}}). The factor model is a prevalent way to
capture the obligors' shared behaviour through a joint common factor, and to 
reduce the dimension of dependence parameters which benefits bond portfolio management. However, one can still find some unrealistic settings on this method such as a constant and linear dependence
structure with thin tails of risk factor distribution embedded.\\

The factor copula model
imposes a dependence structure on common factors and the variables interested. In credit risk measurement, the factor loading represents the sensitivity of the $n$th obligor to the systematic factor. All the
correlations between obligors arise from their dependence on the
common factor. The common factor plays a major role in determining their joint dependence. By applying factor copula model into credit risk modelling, we are able to decompose a latent variable into the systematic and the
idiosyncratic component which are independent. A latent variable usually represents the proxy of firms' assets or liquidation value ({\color{blue}
	\citealp{andersen2004extensions}}). Default is triggered by
company asset values falling below a threshold, representing a
fraction of company debt ({\color{blue}
	\citealp{merton1974pricing}}). In this model, credit risk is measured by a Gaussian random default variable generated from firm asset
value that is latent and modelled by a factor copula framework. The implied firm value from the model ideally projects the default time we desire; that is, a lower firm value is, a shorter default time is. \\

A constant factor loading assumption embedded in a one factor
Gaussian model is inconsistent with the fact that the loading on common factors varies over time, which hampers the measurement of the dependency structures of obligors.
This observation is in fact at the core of research on the
mispricing of structured products ({\color{blue}
\citealp{choros2013valuation}}; {\color{blue}
\citealp{choros2014copula}}).
{\color{blue}\citet{longin2001extreme}} and
{\color{blue}\citeauthor{ang2002asymmetric} (2002b)} argue that a
\textquotedblleft correlation breakdown\textquotedblright \,
structure acts better in the dependence specification. Note that if we
set the factor loading constant, we may underestimate the default
risk as the market turns downward. Our simulation and empirical evidence show that a greater factor loading in market downturn leads to a higher contribution of common factor on firm value.\\

In addition to the specification of factor loading, a
critical and essential part in calculating the portfolio loss
function is recovery rate. According to Table 1, a
state-dependent recovery rate model is suggested since the
recovery rate seems to be subject to the market conditions; that
is, higher in a bull market and lower in a bear market. By closely
looking, one observes a lower average annual recovery rate in the
period 1998 to 2001 (internet bubble) and 2008 to 2009 (US
subprime crisis) compared to the rest of the periods with bullish prospects. It is certain that the recovery rate in
the bull market should not be lower than that in the bear market.
Therefore, the recovery rate is likely to vary with market
conditions, which resembles the behaviour of the default rate. Notice
that the market condition is the unique common factor shared
between recovery rate and default
rate, and causes their time variations. \\
\begin{table}[!ht]
\begin{center}
\begin{tabular}{lrrrrrr}
\hline\hline
  &\multicolumn{6}{c}{Bond}\\
 Year& Sr. Sec. &   Sr. Unsec. & Sr. Sub.  &  Sub.   & Jr. Sub.  &  All Bonds\\
 \hline
1997  &  75.5\%  & 56.1\% &  44.7\% &  33.1\%  & 30.6\% & 48.8\%\\
1998  &  46.8\%  & 39.5\% &  45.0\% &  18.2\%  & 62.0\% & 38.3\%\\
1999  &  36.0\%  & 38.0\% &  26.9\% &  35.6\%  & n.a.   & 33.8\%\\
2000  &  38.6\%  & 24.2\% &  20.8\% &  31.9\%  & 7.0\%  & 25.1\%\\
2001  &  31.7\%  & 21.2\% &  19.8\% &  15.9\%  & 47.0\% & 21.6\%\\
2002  &  50.6\%  & 29.5\% &  21.4\% &  23.4\%  & n.a.   & 29.7\%\\
2003  &  69.2\%  & 41.9\% &  37.2\% &  12.3\%  & n.a.   & 41.2\%\\
2004  &  73.3\%  & 52.1\% &  42.3\% &  94.0\%  & n.a.   & 58.5\%\\
2005  &  71.9\%  & 54.9\% &  32.8\% &  51.3\%  & n.a.   & 56.5\%\\
2006  &  74.6\%  & 55.0\% &  41.4\% &  56.1\%  & n.a.   & 55.0\%\\
2007  &  80.6\%  & 53.7\% &  56.2\% &  n.a.    & n.a.   & 55.1\%\\
2008  &  54.9\%  & 33.2\% &  23.3\% &  23.6\%  & n.a.   & 33.9\%\\
2009  &  37.5\%  & 36.9\% &  22.7\% &  45.3\%  & n.a.   & 33.9\%\\
2010  &  62.5\%  & 51.5\% &  37.5\% &  33.7\%  & n.a.   & 51.8\%\\
2011  &  63.3\%  & 41.3\% &  36.7\% &  35.4\%  & n.a.   & 46.3\%\\
2012  &  51.2\%  & 43.0\% &  33.7\% &  37.3\%  & n.a.   & 44.7\%\\
2013  &  57.7\%  & 43.8\% &  20.7\% &  26.4\%  & n.a.   & 45.6\%\\
\hline\hline
\end{tabular}
\end{center}
\caption{{\bf Annual defaulted corporate bond recoveries}\\
Annual corporate bond recovery rates based on post default trading
price, Moody's 27th annual default study. Note that Sr. Sec., Sr. Unsec., Sr. Sub., Sub., and Jr. Sub. represent senior secured, senior unsecured, senior subordinated, subordinated and junior subordinated, respectively.}\label{recovery}
\end{table}

{\color{blue}\citet{andersen2004extensions}} address that
both default events and recovery rates are driven by a single
factor, but with an independent assumption between default and
recovery rate. There are reasons to doubt this assumption.
{\color{blue}\citet{chen2010macroeconomic}} demonstrates that the
recovery rates are strongly negatively correlated with default
rates (is given as -0.82). As a consequence, the dependence
between them depends on the common factor represented by the state
of macroeconomics. We claim that the common factor (market) governs the default rate and recovery rate simultaneously and creates their association implicitly. One of our purposes is to build a tractable
model that is capable of reflecting the obligors' behaviour in
reacting to the impact from the market. In addition, we show that a systematic risk plays a critical role
in credit measurement and prediction, and contributes more
to a firm's credit risk in a market downturn than in a tranquil
period. In this sense, the factor loading on common factor is conditional on market states. This conditional specification enables risk managers to be alerted regarding risk to the deterioration of the credit conditions when the market
turns
down, which avoids underestimating the default probability.\\

We extend the one factor Gaussian copula model in two ways.
Firstly, to improve the factor loading of
{\color{blue}\citet{andersen2004extensions}} given a two-point
distribution, we apply the state-dependent concept from
{\color{blue}\citet{kim2000stress}} with the specific distributions
to characterise the correlations in hectic or quiet periods, respectively. It
potentially captures two typical features of equity index
distributions: fat tails and a skew to the left. However, for a
two-point distribution setting, it is difficult to decide on the threshold
level of the two-point distribution, and on a time to be chosen
arbitrarily. Secondly, by relaxing the constant recovery rate presumed naively by academia and industry, our state-dependent
recovery rate model permits that the systematic risk factor determines the Loss Given Default (LGD), as suggested by
{\color{blue}\citet{amraoui2012pricing}}. In addition, it
restricts the recovery rate, as a percentage of the notional is
bounded on [0,1] to achieve the tractable and numerically
efficient missions. In summary, we contribute the incorporation of
the state-dependent recovery rate into the conditional factor
copula model, and model them by sharing the unique common factor.
The common factor governs the default rate and recovery rate
simultaneously, and creates their association implicitly. Our
Monte Carlo simulation and empirical evidence
appropriately reflect this feature.\\

We propose four competing default models that have been widely
applied to measure credit risk, and evaluate their relative
performances on the accuracy of forecasting default in the
following year. This comparison, by mapping the various factor
copula models developed in the past and current literature to the
competing models, fosters the discussion on the model performance.
Therefore, to achieve a broader and robust comparison, we group
the factor copula models developed in the literature into four competing
models: (1) The FC model: the standard one-factor
Gaussian copula model with the constant recovery rate
({\color{blue}\citealp{van2007factor}}; {\color{blue}\citealp{rosen2010risk}}). (2) The RFL model: the
one-factor Gaussian copula model with the factor loadings tied to
the state of common factor and the recoveries being assumed constant ({\color{blue}\citealp{kalemanova2007normal}}; {\color{blue}\citealp{chen2014mixed}}). (3) The RR model: standard
one-factor Gaussian copula model but the recoveries being related
to the state of the macroeconomic state
({\color{blue}\citealp{amraoui2008optimal}};
{\color{blue}\citealp{elouerkhaoui2009base}}; {\color{blue}\citealp{amraoui2012pricing}}), and (4) The RRFL
model: a conditional factor
loading specification together with a state-dependent recovery rate, and this is the model what we are developing and contributing to. If further empirical results show its best performance on default prediction, the outstanding performance of our refined RRFL model becomes very clear.\\

In the FC model, we estimate the Spearman's correlation coefficient
between each obligor and common factor and set the recovery rate
as constant. This is a conventional model to measure the capital
requirement in the Basel II accord. By relaxing the constant
correlation in the RFL model, we suggest that the conditional factor loading plays
a significant role in capturing an asymmetric systematic impact
from the market. The RR model uses the method
proposed by {\color{blue}\citet{amraoui2012pricing}} to
investigate the effect of stochastic recovery rate. It allows the
LGD function to be driven by the common factor and the hazard rate, but keeps factor loadings constant. In the RRFL
model, we incorporate the conditional factor loading into
state-dependent recovery rate and model them by sharing the unique
common factor. To evaluate whether these two specifications carry
significant improvements to the default prediction, we use the data
set of daily stock indices of the S\&P 500  to represent the market (common factor) and the respective stock prices
of the default companies for the period of 5 years before
the default year from the Datastream database. 

Our default data analysis contains 2008 and 2009, as collected by
Moody's report. We use Moody's Ultimate Recovery Database (URD)
which is the ultimate payoff that obligors can obtain when the default
emerges from bankruptcy or is liquidated instead of the
post-default trading price as proposed by 
{\color{blue}\citet{carty1998bankrupt}}. They examine whether the
trading price represents a rational forecast of actual recovery,
and find that it is not a rational estimation of actual
recovery. For this period, we employ a state-dependent concept in
order to capture an asymmetric impact from the common risk factor.
As a result, we achieve the goal that both conditional factor loading and
state-dependent recovery rates improve the calibration of our default
prediction. The conventional factor copula underestimates the
impact of systematic risk and portfolio credit loss when the market is in downturn. We find that the incorporation of
factor loading into the state-dependent recovery rate improves the accuracy of the default prediction. This
result is coherent with the goal of Basel III, which emphasises the role of systematic risk on overall
financial stability and systemic risk. In our later empirical analysis, we concentrate on the senior unsecured bond, since there is a rich data source available.\\

The remainder of the study is organised as follows. Section 2
describes the goal of Basel III. We present a general framework
and the standard one-factor Copula in section 3. Besides, we
extend the standard one-factor Copula by using the conditional factor
loading and the state-dependent recovery model. Section 4 describes
the data set. In the section 5, we offer empirical evidence.
Section 6
presents the conclusion.\\

\section{Systematic risk in Basel III}

As highlighted by Basel III, systemic risk is crucial in financial markets from several aspects. First, a bank can trigger a shock throughout a system and spill over to its counterparties ({\color{blue}\citealp{drehmann2013measuring}}).
Secondly, procyclicality could destabilise the whole
systemic risk ({\color{blue}\citealp{basel2009strengthening}}). The borrowers hardly fund more as their collateral assets have depreciated caused by weak economic conditions. Third, since Basel II focused on minimising the default probability of individuals, this accord failed to guarantee a stable financial system due to a lack of concern for systemic risk. Therefore, a new Basel accord is expected to emphasize its role.\\

Systematic factor is one of the important drivers of systemic risk and probably constituting a serious threat to systemic fragility ({\color{blue} \citealp{schwerter2011basel}}; {\color{blue} \citealp{uhde2010securitization}}). {\color{blue}\citet{tarashev2010attributing}} also distinguish between systemic risk and systematic risk. The former refers to the risk that impedes the financial system, while the latter refers to the commonality in risk exposures of financial institutions. Their model assumes that systemic risk can have systematic and idiosyncratic components. It is understandable that systemic risk is heightened by systematic risk. A bank is characterised as one of systemically important (too-big-to-fail) financial institutions, its default would lead to a dramatic impact on systemic risk. This is the very reason what Basel III attempts to regulate and prevent. Through our paper, our model proposes that the contribution of systematic risk is higher than that of idiosyncratic component, and this dominance is characterised by a higher factor loading on systematic risk during a market downturn. We, therefore, see the contribution of systematic risk on credit risk varies with time and market conditions. In this regard, one shall concern on the interconnection between credit risk and market risk. It is worth noting that the points mentioned above determine the sufficiency of capital requirement in the banking industry.\\

To obtain the sufficient capital requirements, recovery rate is one of determinant variables in credit risk estimation. A real observation is that in a recession period, recovery rates tend to decrease while default rates tend to rise. As such, increasing capital requirement under this condition seems necessary. Most early academic studies on credit risk assume that recovery rates are deterministic ({\color{blue} \citealp{schonbucher2001factor}}; {\color{blue} \citealp{rosen2010risk}}), or they are stochastic but independent from default probabilities ({\color{blue} \citealp{jarrow1997markov}}; {\color{blue} \citealp{andersen2004extensions}}). Neglecting the nature of stochastic in recovery rate and the interdependence between recovery rates and default rates result in a biased credit risk estimation ({\color{blue} \citealp{altman2005link}}).\\

To be close to the spirit of Basel III, our study extends the existing literature into two dimensions. First, we highlight that systematic risk is a predominant factor in a recession period, and proceed a relative contribution analysis to measure the proportional contribution from a systematic risk in comparison with that from an idiosyncratic component. Second, we propose a methodology in which recovery rates and default rates are correlated by sharing a unique factor, and both are state-dependent. Our model design, the simulation and empirical results provide a bundle of justifications for the goals of Basel III.

\section{Methodology}
\subsection{Default modelling}

Recognising the importance of systematic risk, one-factor Gaussian models have been considered an important tool
underlying the internal ratings based approach ({\color{blue}
\citealp{crouhy2000comparative}}; {\color{blue}
\citealp{pykhtin2002credit}}; {\color{blue}
\citealp{frey2003dependent}}) and used to price CDOs
({\color{blue} \citealp{hull2004valuation}}; {\color{blue}
\citealp{andersen2004extensions}}; {\color{blue}
\citealp{choros2013valuation}}). It reduces the number of
correlations being estimated from $\frac{N(N-1)}{2}$ by a
multivariate Gaussian Model to $N$ which represents the number of assets. Specifically, we use
a non-standardised Gaussian model to represent the deteriorating
market condition by presuming a negative mean value together with a higher volatility.
The model is based on decomposing a latent variable $U_{i}$ for obligor $i$ into the systematic factor $Z$ and the idiosyncratic component $\varepsilon_{it}$:\\

\begin{equation}
U_{i}=\alpha_{i}Z+\sqrt{1-\alpha_{i}^{2}}\varepsilon_{i}
\hspace{0.5cm}i=1,\ldots,N \end{equation}\\

\noindent where $-1\leq\alpha_i\leq1$. Suppose that $Z\sim
\textsf{N}(\mu, \sigma^{2})$ and $\varepsilon_{i}$ have zero-mean
unit-variance distributions. In a Gaussian content, $Z$ and
$\varepsilon_{i}$ are orthogonal and $\varepsilon_{i}$ are
mutually uncorrelated. The distribution of vector $U$ can be
described by a copula
function which joins two marginals, $Z$ and $\varepsilon_{i}$. The correlation coefficient $\rho_{ij}$ between $U_i$ and $U_j$ can be described by their $\alpha_i$ and $\alpha_j$:\\

\begin{equation}
\rho_{ij}=
\frac{\alpha_i\alpha_j\sigma^{2}}{\sqrt{\alpha_i^{2}(\sigma^{2}-1)+1}\sqrt{\alpha_j^{2}(\sigma^{2}-1)+1}}
\end{equation}\\

\noindent where $\sigma_i=\sqrt{\alpha_i^{2}(\sigma^{2}-1)+1},
\sigma_j=\sqrt{\alpha_j^{2}(\sigma^{2}-1)+1}$. As a consequence,
the number of correlations describing the dependency structure is
reduced in size since only $N$ parameters ${\alpha_i:
i=1,\ldots,N}$ need to be estimated.
We express the covariance matrices between $U_i$ and $U_j$ under a factor model,\\
\begin{equation}
\Sigma_{ij}=\sigma_i^{2}\sigma_j^{2}\left(\begin{array}[c]{cc}
1 & \rho_{ij}\\
\rho_{ji} & 1 \end{array}\right)
\end{equation}\\

\noindent The one-factor Gaussian copula model we consider is used
to model the default indicators to time $t$, $\mathbf{I}\left\{\tau_{i}\leq t\right\}$, by projecting $U_i$
into $\tau_i$. $U_i$ here can be viewed as the proxies for firm
asset and liquidation value ({\color{blue}
\citealp{andersen2004extensions}}). In this regard, the lower
asset value of firm the shorter time to default, $\tau_i$. More
precisely, $U_i \leq F^{-1}\{P_i(t)\}$ leads to $\tau_i\leq t$, where
$P_i(t)$ is a hazard rate and marginal probability that obligor $i$
defaults before $t$, and $F^{-1}(\cdot)$ donates the inverse cdf of any distribution. The default indicator then can be written as\\

\begin{equation}
\mathbf{I}\left\{\tau_{i}\leq t\right\} =
\mathbf{I}\left[U_{i}\leq F^{-1}\{P_i(t)\}\right]
\end{equation}\\

Given the LGD for each $i$, $G_i, i=1,\ldots,
N$, we aggregate them as total portfolio loss, $L$, as following,\\

\begin{equation}
L= \displaystyle\sum_{i = 1}^{N}G_{i}\mathbf{I}\left\{\tau_{i}\leq
t\right\}= \displaystyle\sum_{i = 1}^{N}
G_i\mathbf{I}\left[U_{i}\leq F^{-1}\{P_i(t)\}\right]
\end{equation}\\

\subsection{Conditional default model}

In accordance with the spirit of Basel III, the systematic latent factor, $Z$, representing the general economic
condition that characterises the systematic credit risk influences the default probability $P_i(t)$ and the
recovery rate $R_i=1-G_i$. So given $Z$, one may write the
conditional default probability $P_i(Z|S=H,Q)$
and conditional LGD, $G_i(Z|S=H,Q)$ as a function of $Z$, and it is state-dependent, $\mbox{S}\in\{\mbox{H,Q}\}$. H, and Q represent the hectic and quiet periods, respectively.\\

A higher factor loading, $\alpha_{i}$ in equation (1) has been observed in hectic periods ({\color{blue}
\citealp{longin2001extreme}}; {\color{blue}\citeauthor{ang2002international}
2002a}; {\color{blue}\citeauthor{ang2002asymmetric} 2002b}). This
observation can be modelled by a regime-switching mechanism, requiring a
globally valid time series structure for $\alpha_i$ from $t$.
Avoiding such a possible too rigid structure, we assume the two
asset returns, $Z$ common factor proxied by USD S\&P 500, $U_i$ (firm stock price) have a mixture of bivariate normal distribution
(See Appendix A) to obtain the estimation of  $\alpha^{H}_{i}$ and  $\alpha^{Q}_{i}$. Given the
conditional factor loading, $\alpha^{H}_{i},
\alpha^{Q}_{i}$, the conditional default model is defined as following,\\

\begin{equation}
U_{i}|_{\mbox{S=H}}=\alpha^{H}_{i}Z+\sqrt{1-(\alpha^{H}_{i})^{2}}\varepsilon_{i}
\end{equation}\\

\begin{equation}
U_{i}|_{\mbox{S=Q}}=\alpha^{Q}_{i}Z+\sqrt{1-(\alpha^{Q}_{i})^{2}}\varepsilon_{i}
\end{equation}\\

\noindent Therefore, the state-dependent conditional default probability can be denoted by\\

\begin{equation}
P(\tau_i<t|\mbox{S})=F\left[{\frac{F^{-1}\{P_i(t)\}-\alpha^S_iZ}{\sqrt{1-(\alpha^S_i)^2}}}\right]=P_i(Z|\mbox{S})\hspace{0.5cm}\mbox{S}\in\{\mbox{H,Q}\}
\end{equation}\\

\noindent Given $P_i(t)$, if the factor loadings
in hectic periods are greater than ones in quiet days, say $\alpha^{H}
> \alpha^{Q}$, and if the index return of S\&P 500 is negative in a bad
market condition, both conditions will result in a higher conditional default
probability in equation (8). From equation (8), the systematic
risk, $Z$, and the corresponding factor loading govern the
conditional default probability, which is consistent with
empirical findings ({\color{blue}\citealp{andersen2004extensions}}; 
{\color{blue}\citealp{bonti2006credit}}). It is worth pointing out that $\alpha^{S}_{i}$ is state-dependent instead of a
constant setting in previous literature
({\color{blue}\citealp{andersen2004extensions}}; 
{\color{blue}\citealp{amraoui2012pricing}}).
{\color{blue}\citeauthor{ang2002asymmetric} (2002b)} set
probability of both regimes equally ($w=0.5$), instead, we
estimate it from the historical data of the S\&P 500 Index return proxied for systematic risk, $Z$, P(S=H)=$\omega$, P(S=Q)=$1-\omega$ by
Expectation-Maximization (EM) algorithm. \\

Likewise, the recovery rates can be designed in this way by incorporating market condition as a main driver across different states. Based on the finding of
{\color{blue}\citet{das2009hedging}}, recovery rates are
negatively correlated with probabilities of defaults and driven by market condition. By relaxing
constant recovery rates, we follow
{\color{blue}\citet{amraoui2012pricing}} to connect recovery rates
and default events via a common factor, but extend their model to a
conditional or state-dependent framework. The recovery rate is
governed by the state of economy, in addition, we
incorporate a conditional correlation structure, $\alpha^S_i$, into stochastic recovery rate model, and set $R_{i}(Z|S=H,Q)$, of obligor $i$, in relation
to the common factor $Z$ and the marginal default probability
$P_i$. The state-dependent recovery rate is expressed as,\\

\begin{equation}
G_{i}(Z|\mbox{S=H})=(1-\overline{R_i})\frac{F\left[ \{F^{-1}\left(
\overline{P}_{i}\right) -\alpha^{H}_{i}Z \}
/\sqrt{1-(\alpha^{H}_{i})^2}\right] }{F \left[ \{
F^{-1}\left(P_{i}\right) -\alpha^{H}_{i}Z\} /\sqrt{%
1-(\alpha^{H}_{i})^2}\right]}
\end{equation}\\

\begin{equation}
G_{i}(Z|\mbox{S=Q})=(1-\overline{R_i})\frac{F\left[ \{
F^{-1}\left( \overline{P}_{i}\right) -\alpha^{Q}_{i}Z \}
/\sqrt{1-(\alpha^{Q}_{i})^{2}}\right] }{F
\left[ \{F^{-1}\left(P_{i}\right) -\alpha^{Q}_{i}Z\} /\sqrt{%
1-(\alpha^{Q}_{i})^{2}}\right]}
\end{equation}\\

\noindent In equation (9, 10), $0\leq \bar{R_i}\leq R_{i} \leq 1$
which means a downward shift of $\bar{R_i}$ to $R_{i}$, so that
$\bar{R_i}=R_{i}-\upsilon$ and $R_{i}\geq \upsilon > 0$. $\upsilon$ is size
of downward shift. By assuming that expected loss in name $i$
remains unchanged, we set $(1-R_i)P_i=(1-\bar{R}_i)\bar{P}_i$.
Please see the proof in A.1 in
{\color{blue}\citet{amraoui2012pricing}}. $F(\cdot)$ denotes any
distribution and $\overline{P}_{i}$ is the adjusted default
probability calibrated proposed by
{\color{blue}\citet{amraoui2008optimal}}. The LGD function,
$G_i(Z|\mbox{S=H,Q})$ essentially can be obtained according to
formula (9,10). Numerous studies show that recoveries decline
during recessions ({\color{blue} \citealp{altman2005link}}; 
{\color{blue} \citealp{bruche2010recovery}}). Consistent with the
spirit of equation (6,7),  we design $\alpha^H$, $\alpha^Q$, the
factor loading in equation (9,10) are therefore conditional and
state-dependent. Moreover, a partial derivative of LGD function
with respect to $Z$ is less than zero proved by property 3.2 in
{\color{blue}\citet{amraoui2008optimal}}, which means that
$G_i(Z|\mbox{S=H,Q})$ is decreasing in $Z$. By assuming $\alpha^H
> \alpha^Q$, which means that a higher factor loading that is usually accompanied by a bad market condition on $Z$ tends to increase LGD. The magnitude of LGD is not only influenced by $Z$ but also sensitive to the factor loading under $Z$; this is what we point out and contribute to the literature. In addition, recovery rates are
also linked to the probability of default and they are negatively
correlated (see {\color{blue} \citealp{altman2005link}}; 
{\color{blue} \citealp{khieu2012determinants}}). With
$Z$, $P_i$ and the estimated conditional factor loading
$\alpha^H$, $\alpha^Q$, we obtain the state-dependent recovery
rate, $R_i(Z|\mbox{S=H,Q})$, and state-dependent LGD, $G_i(Z|\mbox{S=H,Q})=1-R_i(Z|\mbox{S=H,Q})$.

With these two specifications, the conditional default probability
$P_i(Z|\mbox{S=H,Q})$ and conditional LGD, $G_i(Z|\mbox{S=H,Q})$,
conditional expected loss,
therefore, is\\

\begin{equation}
\textsf{E}(L_i|Z)=\omega
G_i(Z|\mbox{S=H})P_i(Z|\mbox{S=H})+(1-\omega)G_i(Z|\mbox{S=Q})P_i(Z|\mbox{S=Q})
\end{equation}\\
\noindent where $\omega=\mbox{P(S=H)}$ and $1-\omega=\mbox{P(S=Q)}$. H and Q represent the hectic and quiet periods, respectively.

\subsection{Monte Carlo simulation}

In this section, we investigate the performance of default
prediction by establishing a simulation of realistic scenarios.
The default probability and recovery rate function are governed by
systematic factors generated from different regimes. Indeed, they
are crucial elements in evaluating the accuracy of the default
prediction. Our interest is to see whether the design of
conditional factor loadings and state-dependent recovery rates contribute to the
default prediction.

\subsubsection{One-factor non-standardized Gaussian copula}

We simulate one-factor non-standardised Gaussian copula subject to
different states. As described in equation (6) and (7), we
generate systematic factor $Z$ by non-standardised Gaussian
distribution with different volatilities and independent
$\varepsilon_i's$ . To reflect the nature of distinct
variations exhibited in different market conditions.\\

Through a mixture bivariate distribution setting in Appendix A,
the conditional factor loadings, $\alpha^H_i$ and $\alpha^Q_i$ are
derived, in the one-factor non-standardised Gaussian copula model.
We estimate them from the daily stock returns of S\&P 500 and of
collected default companies during the crisis (2008-2009) period.
The five-year period prior to the crisis period is the estimation
period for the conditional factor loadings. The return of S\&P 500
Index represented as a systematic factor, $Z$, is presumed to
distribute as $\textsf{N}(-0.03, 3.05)$ estimated in 2008 and
2009, while $\varepsilon_i\sim \textsf{N}(0,1)$ represents
idiosyncratic risk. $Z$ and $\varepsilon_i$ are generated 1000
scenarios, respectively. Given any one of generated systematic
factor random variables, $Z$, and using Bayes' rule, we calculate
the conditional probability that date $t$ belonged to the hectic
is $\pi(Z=z)$ by using its counterpart,
unconditional probability $\omega$, as a formula (12). \\

\begin{equation}
\mbox{P}(S=H|Z=z)=\pi(Z=z)=\frac{\omega
\varphi(z|\theta^H)}{(1-\omega)\varphi(z|\theta^Q)+\omega
\varphi(z|\theta^H)}
\end{equation}\\

\noindent $\theta^{H}, \theta^{Q}$ represent the parameters of
distribution in the hectic (H) and the quiet (Q) period. $\varphi(\cdot)$
is a normal distribution. Plugging $\alpha^H_i, \alpha^Q_i$ shared
with the same simulated $Z$ random variables, conditional $U_i|$S
is generated as developed in equation(6, 7). These simulated
random variables together with the published hazard rates $h_i(t)$
ideally produce the simulated default times.

\subsubsection{Default time}

Projecting $U_i$ simulated from section 3.3.1 to default time, $\tau_i$, stated
in equation(4) provides the clue as to whether the firm defaults before
time. We set $t=1$, represents the time interval of 1 year, so
that $\tau_i<1$ is referred to a default event in the $i$th
obligor. The hazard rate $h_i$ is the probability of
occurrence of the default event within one year. $\tau_i$ is
referred to default time of $i$th obligor. More precisely, the
expected value of $\textsf{E}[\mbox{I} (\tau_i<1)]$ is
$\mbox{P}(\tau_i<1)$ or named as $P_i$, see
{\color{blue}\citet{franke2015statistics}} Chapter 22, that can be
connected to the firm's stock return or firm's value, $U_i$ leads
to $P_i=\textsf{E}[\mbox{I}\{U_i<F_i^{-1}(P_i)\}]$ where $F_i$
denote the cdf of $U_i$. By applying generated $U_i$ from the
conditional factor model into the definition of the survival rate, we
have generated default time, $\tau_i$, derived from
$1-\mbox{exp}(-P_i\tau_i)=F(U_i)$
({\color{blue} \citealp{hull2006options}}). To keep on the state-dependent environment, the conditional default time for each obligor is generated by formula (13).\\

\begin{equation} \tau_{i}|\mbox{S} =
-\frac{\mbox{log}\{ 1 -F( U_{i}|_{\mbox{S}})\}}{P_i}
\end{equation}\\

\noindent where $P_i$ is the hazard rate or marginal probability
that obligor $i$ will default during the first year, conditional on
no earlier default, and is obtained from Moody's report. It is the
cumulative of default rates during the first year. Equation (13) states that 
$U_{i}|\mbox{S}$ becomes larger, $\tau_{i}|\mbox{S}$ will become
longer. The larger $U_i$ reduces the tendency of default and
postpones the default time, $\tau_{i}|\mbox{S}$.\\

\subsubsection{State-dependent recovery rate simulation}
In the third step, we consider a more realistic situation by
simulating recovery rates as described in our settings. The
adjusted default probability $\bar{P_i}$ is calibrated by using
hazard rate $P_i$ from Moody's report. $\bar{R_i}$ is a lower
bound for state-dependent recovery rates [0,1], therefore, we set
$\bar{R_i}=0$ in the simplest case. With $\alpha^H_i, \alpha^Q_i,
Z, \bar{P_i}$, the simulated state-dependent recovery rates are obtained by
formula (9, 10).

\subsubsection{Loss function}
By changing scenarios to quiet and hectic states, we assume the exposure of each obligor is 100 million and generate the expected
loss
under the given scenarios corresponding to formula (11). \\

\begin{equation}
 \textsf{E}(L_i|Z)=\pi(Z=z)G_i(Z|\mbox{S=H})P_i(Z|\mbox{S=H})+(1-\pi(Z=z))G_i(Z|\mbox{S=Q})P_i(Z|\mbox{S=Q})
\end{equation}\\

\noindent Given the simulated Z random variables, conditional probability $\pi(Z=z)$ naturally provides better information than unconditional probability $\omega$ does. By the given formula (14), we
compare the theoretical loss amounts across four models with the
realised loss values, and evaluate the performance of the default prediction by the mean of square error.\\

\subsubsection{Absolute error}

In step 5, the performance of the competing models: FC, RFC, RR,
RRFC are evaluated here to decide which one is the best in
predicting the default for the following year. Absolute Error (AE)
here is linked to the prediction performance and is defined as\\

\begin{equation}
\mbox{AE}=(\mbox{actual portfolio loss - expected portfolio loss})
\end{equation}\\

\noindent where actual portfolio loss is from Moody's report.
Expected loss is estimated from equation (14), whereas in an
unconditional default model, it is
computed from formula (5). For each competing model, we generate 1000 scenarios, then, the mean of absolute error referred as MAE is calculated. 
One can expect that the best one is entailed on the minimum AE and MAE as well. 

\section{Data}

We use the list of default companies for 2008 through to 2009
published by Moody's annual report since this is a rich available data
source. In total, we obtained 341 defaults with
corporate bond recovery rates from Moody's URD covering the period
from 1987 to 2007. We focus on senior unsecured bonds because of their wide use in
financial contracts, regulatory rules, and the risk of measuring for
assets under the standardised approach of Basel II ({\color{blue}
\citealp{pagratis2009modeling}}). We also collected the credit rating
of obligors from Moody's report in order to measure the hazard
rate. Although there are 94 and 247 default firms in 2008 and
2009, the observations were reduced due to missing stock prices and
credit rates of obligors' bonds. If there was a lack of stock
prices of default subsidiary companies, we used stock prices of
parent companies instead. In all cases, 31 and 62 sampling firms were collected in 2008 and 2009, respectively.\\

To estimate the conditional factor loadings of sampling firms, we collect the daily USD
S\&P 500 return and the respective stock return of the default companies for
a 5-year period prior to the default year from the Datastream
database. USD S\&P 500 Index here simply represents the common systematic risk. By assuming a mixture of bivariate normal distribution, we estimate the parameters including factor loadings by EM algorithm. Table 2
presents the results of EM algorithm.\\
\begin{table}[!ht]
\begin{center}
\begin{tabular}{lrrr}
\hline\hline
  Model & Probability & Mean & STD\\
\hline
  Period & 2003-2007\\
\hline
Unconditional (one normal) & 100.00\% & 0.03\% & 0.77\%\\
Conditional on quiet & 58.68\% & 0.10\% & 0.43\% \\
Conditional on hectic & 41.32\% & -0.08\% & 1.07\%\\
\hline
 Period & 2004-2008\\
\hline
Unconditional (one normal) & 100.00\% & 0.03\% & 0.83\%\\
Conditional on quiet & 56.77\% & 0.10\% & 0.38\% \\
Conditional on hectic & 43.23\% & -0.06\% & 1.17\%\\
\hline \hline
\end{tabular}
\end{center}
\caption{{\bf Estimate mixture of normal distribution by employing
an
EM algorithm}\\STD represent standard deviation\\
}\label{EM}
\end{table}

As presented in Table 2, the volatility of the hectic distribution
is larger than that of the quiet distribution, and the mean of the
hectic distribution is smaller than that of the quiet
distribution, reflecting the fat tails and a skew
to the right which are consistent with
{\color{blue}\citet{kim2000stress}}.\\
\section{Empirical result}
\subsection{Conditional factor loading estimation}

Figure 1 and 2 shows that the
majority of correlation coefficients or called factor loadings in factor copula model  during the hectic period is
higher than in the quiet period. The proposed correlation structure leads to more accurate and realistic implementations, and to avoid the underestimation of factor loading in a hectic period or the overestimation in a quiet period. These ideas
are well known in statistics and have already been applied to
financial questions ({\color{blue}\citeauthor{ang2002asymmetric}
2002b}; 
{\color{blue}\citealp{patton2004out}}).\\

\begin{figure}[!ht]
       \begin{center}
			\includegraphics[scale=0.6]{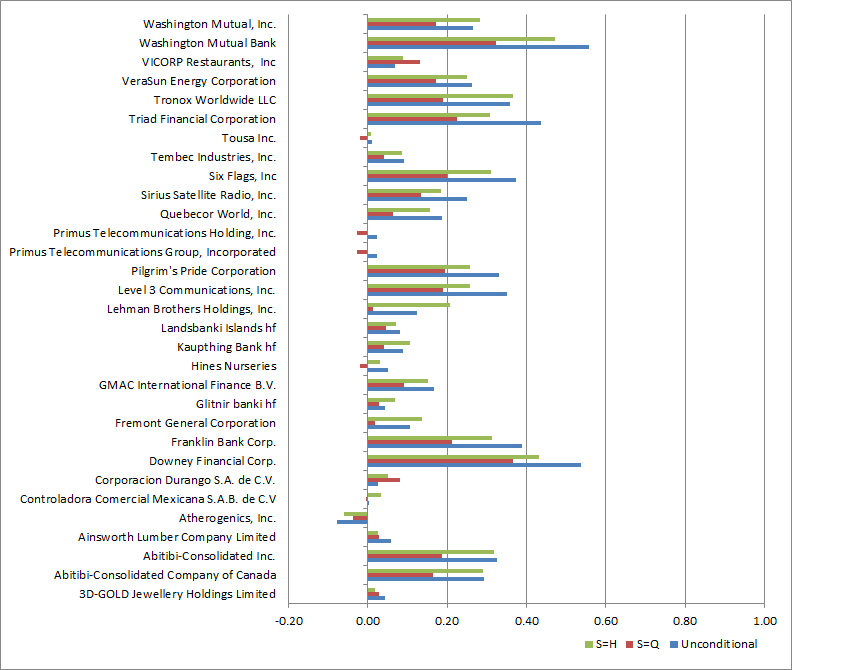}
		\end{center}
			\caption{{\bf Conditional and unconditional factor loading comparision in 2008}\\
				The estimation of Conditional and Unconditional Factor Loading between S\&P 500 and default companies in 2008.} \label{figure1}
\end{figure}

\begin{figure}[!ht]
 \begin{center}
 	\includegraphics[scale=0.6]{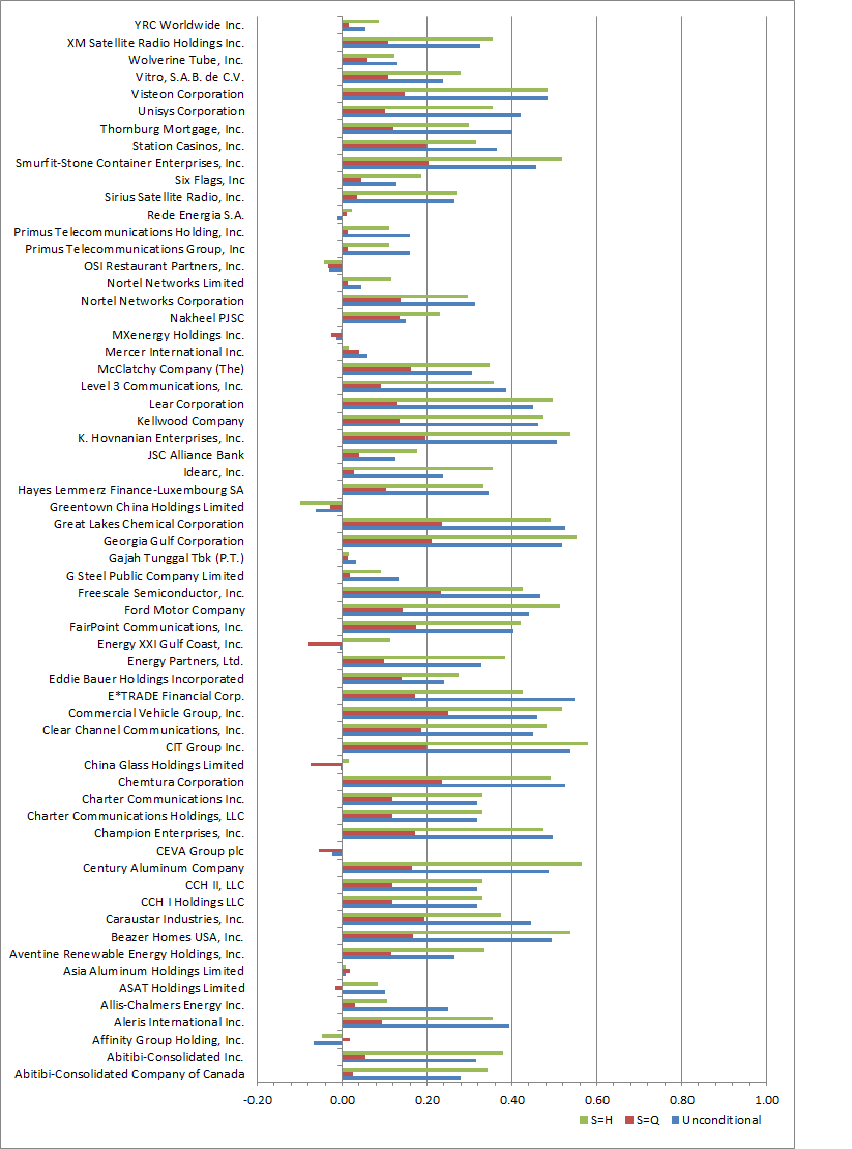}
 \end{center}
 \caption{{\bf Conditional and unconditional factor loading comparision in 2009}\\
 	The estimation of Conditional and Unconditional Factor Loading between S\&P 500 and default companies in 2009.} \label{figure2}
\end{figure}


In our approach, we consider this asymmetric correlation structure under
real market conditions to implement the conditional default model
developed in Section 3.2. As shown in Figure 1 and 2, the factor loadings $\alpha_i$ in state H are higher than those in state Q. As factor loadings get higher in state H, the correlation coefficient $\rho_{ij}$ between firm $i$ and $j$ defined in equation (2) is expected to increase in this market condition. Therefore, obligors tend to comove more
closely during hectic periods than during quiet periods.\\

\subsection{State-dependant recovery rate estimation}

To demonstrate the impact of market conditions measured by Z on
the state-dependent recovery rate, in Figure 3 we depict the
relationship between the state-dependent recovery rate and the
S\&P 500 (the proxy for systematic factor Z) in blue
\textquoteleft *\textquoteright, which developed in section 3.2.
One can observe that the effect of the systematic factor on the
recovery rate is positive, the recovery rate gets higher as $Z$ grows. Since the slope of this curve is
influenced by estimated $\alpha^H_i, \alpha^Q_i$ corresponding to
formula (9, 10), the slopes behave differently in the four panels but keep positive
monotonically. We also depict the stochastic recovery rates in
red \textquoteleft +\textquoteright\, estimated and simulated through
{\color{blue}\citet{amraoui2012pricing}} model, in comparison with blue \textquoteleft *\textquoteright, simulated from our model. Taking (c) E*TRADE as an example, we observe that compared with the simulated recovery rates based on equation (9) and (10), those generated from {\color{blue}\citet{amraoui2012pricing}}, by assuming constant factor loadings, tend to produce higher recovery rates in the market downturn and lower ones in the booming market. This evidence suggests that the recovery rate may be overestimated in a bearish market but underestimated in a bullish market if the constant factor loading is assumed. As a consequence, an underestimation of credit loss in a bearish market but an overestimation in a bullish market are highly possible. Similarly, the evidence from (a) Glitnir
banki (b) Lehman Brothers Holdings, Inc. and (d) Idearc, Inc. are comparable and consistent. Note that the impact of the systematic factor on recovery rate seems nonlinear, it is higher in the market downturn but relatively milder in the booming market, and the marginal slope decreases abruptly when the index return
decreases, whereas the marginal slope decelerates when the index
return becomes positive. This simulation result is in accordance with Moody's
report in Table 1. From 2004 to 2006, the annual recovery rate of
senior unsecured bond increases slowly. When the crisis started in
August 2007, the recovery rate drops dramatically. By capturing the
correlation structure, $\alpha^H > \alpha^Q$, as shown in (a), (c)
and (d), we find this asymmetric pattern which is more consistent with
the reality.\\

Having the simulated recovery rates from equations (9, 10), we are more interested in the relation between it and conditional default probability from equation (8). As can been seen in Figure 4, the simulation result shows
the downward trend between default probability and recovery rate consisted with
{\color{blue}\citet{altman2005link}} and
{\color{blue}\citet{das2009hedging}}. It shows that the common factor governs the default rate and recovery rate simultaneously and creates their negative association implicitly. 
{\color{blue}\citet{altman2005link}} find that permitting a 
dependence between default rates and recovery rates increases
around 29\% in the Value at Risk compared with a model that assumes no
dependence between default rates and recovery rates.\\

\begin{figure}[!ht]
    \begin{subfigure}{0.5\textwidth}
       \begin{subfigure}{\textwidth}
          \renewcommand\thesubfigure{\alph{subfigure}}
          \centering
          \includegraphics[scale=0.45]{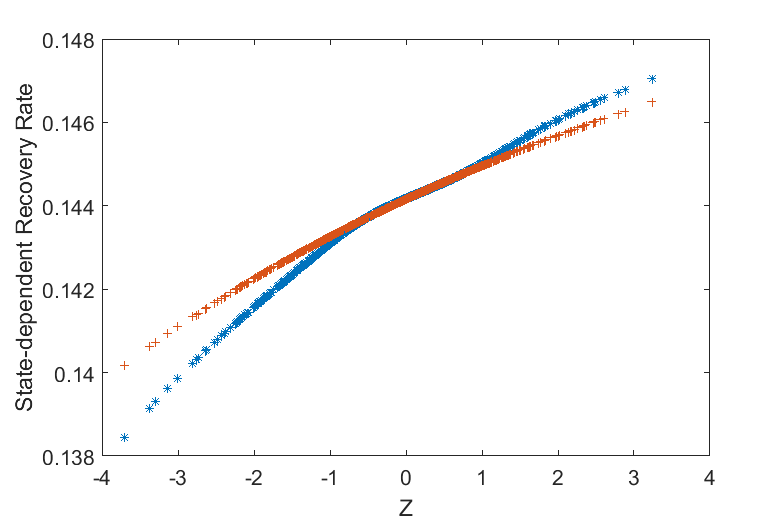}
          \caption{Glitnir banki:\\
          $\alpha=0.044, \alpha^{Q}=0.029, \alpha^{H}=0.067$}
       \end{subfigure}
       \begin{subfigure}{\textwidth}
          \renewcommand\thesubfigure{\alph{subfigure}}
          \centering
          \includegraphics[scale=0.45]{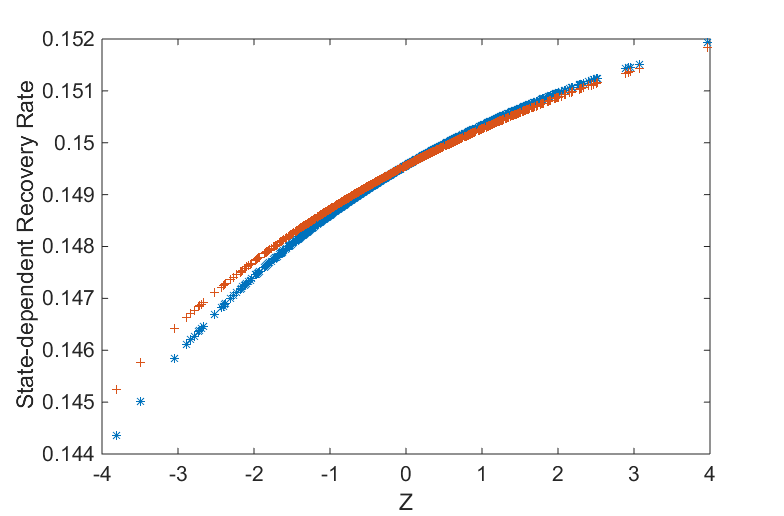}
          \caption{Lehman Bro.:\\
          $\alpha=0.125, \alpha^{Q}=0.013, \alpha^{H}=0.208$}
       \end{subfigure}
    \end{subfigure}
    \begin{subfigure}{0.5\textwidth}
       \begin{subfigure}{\textwidth}
          \renewcommand\thesubfigure{\alph{subfigure}}
          \centering
          \includegraphics[scale=0.45]{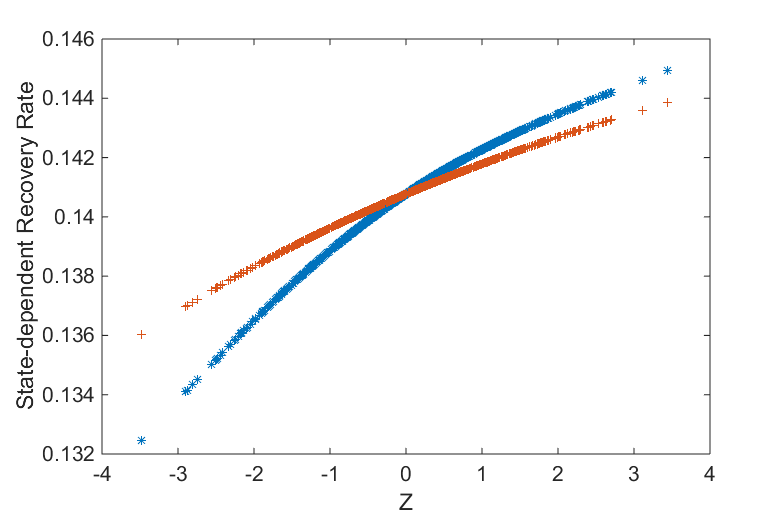}
          \caption{E*TRADE:\\
          	$\alpha=0.548, \alpha^{Q}=0.172,\alpha^{H}=0.426$}
       \end{subfigure}
       \begin{subfigure}{\textwidth}
          \renewcommand\thesubfigure{\alph{subfigure}}
          \centering
          \includegraphics[scale=0.45]{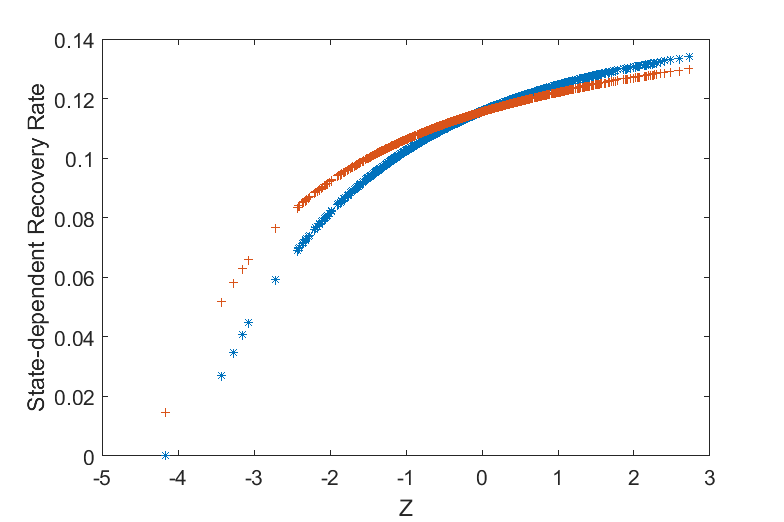}
          \caption{Idearc, Inc.:\\
          	$\alpha=0.237, \alpha^{Q}=0.028,\alpha^{H}=0.356$}
       \end{subfigure}
    \end{subfigure}
  \caption{{\bf The relationship between state-dependent recovery rate and index return of S\&P 500,
  $Z$.}\\
           Panel (a) and (b), \textquoteleft *\textquoteright\, in blue illustrates the pattern of state-dependent recovery rate of Glitnir banki and Lehman Brothers Holdings,
           Inc. which incorporate conditional factor loading in 2008. \textquoteleft +\textquoteright\, in red plots the recoveries proposed by {\color{blue}\citet{amraoui2012pricing}}. In
           panel (c) and (d), E*TRADE Financial Corp. and  Idearc, Inc. in 2009.
          } \label{figure3}
\end{figure}

\begin{figure}[!ht]
  \begin{center}
  \includegraphics[scale=0.4]{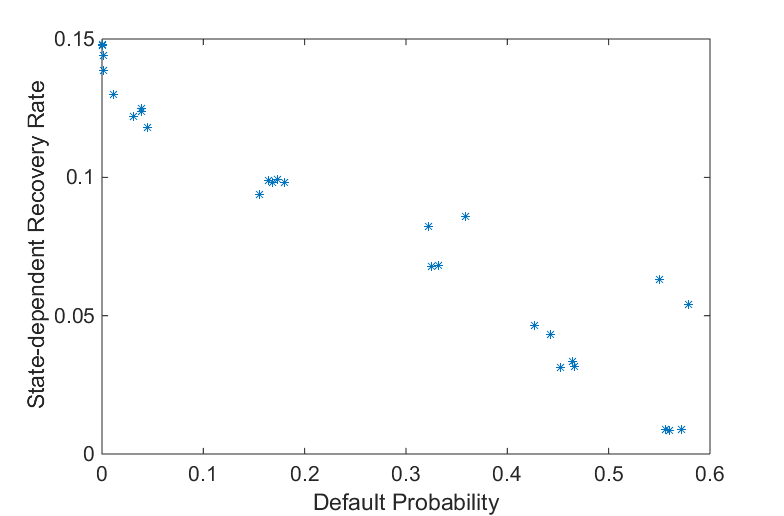}~~~~~~~
  \includegraphics[scale=0.4]{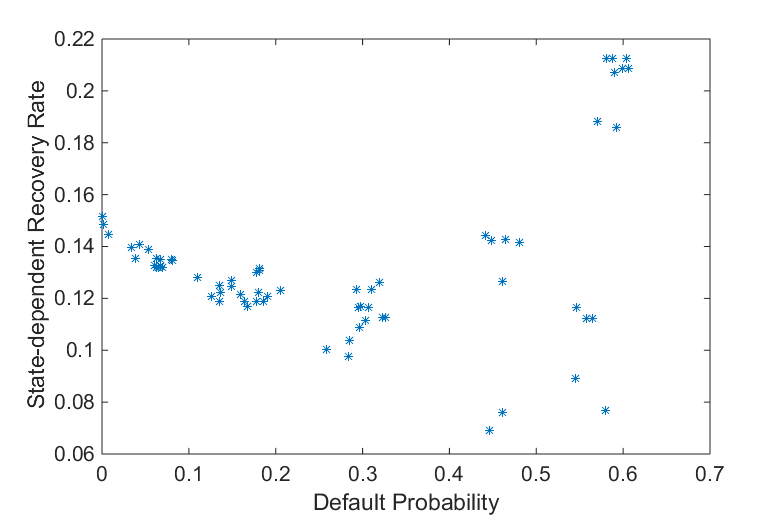} \\
  \parbox[t]{0.5\textwidth}{
  \centerline{ (a) 2008 }
  }\hfill\quad
  \parbox[t]{0.45\textwidth}{
  \centerline{ (b) 2009 }
  }
  \end{center}
    \caption{{\bf The relationship between state-dependent recovery rates and default
    probabilities}\\
           By simulating $Z \sim \textsf{N}(-0.03, 3.05)$, it plots the relationship between the state-dependent recovery rate and default probabilities, given the conditional factor
           loading. By simulating 1000 observations, we estimate
           the default probabilities and state-dependent recovery rate from formula (8) and (9,10).
 } \label{figure4}
\end{figure}

\subsection{Empirical results of absolute errors}
To gauge the conditional factor loading and state-dependent
recovery rate approaches for default prediction, we propose four
models: (1) The FC model: the
standard one-factor Gaussian copula model with the constant
recovery rate developed by {\color{blue}\citet{van2007factor}} and {\color{blue}\citet{rosen2010risk}} (2) The RFL model: the
one-factor Gaussian copula model with the factor loadings tied to
the state of common factor and the recoveries being assumed as constant proposed by {\color{blue}\citet{kalemanova2007normal}} and {\color{blue}\citet{chen2014mixed}}. (3) The RR model:
standard one-factor Gaussian copula model but the recoveries being
related to the state of the macroeconomic state ({\color{blue}\citealp{amraoui2008optimal}}; 
{\color{blue}\citealp{elouerkhaoui2009base}}; {\color{blue}\citealp{amraoui2012pricing}}), and (4) The RRFL
model: a conditional factor
loading specification together with a state-dependent recovery rate. We address the question of whether the two specifications,
conditional factor loading and the state-dependent recovery rate model
are meaningful and significant in explaining the gap between expected and
practical loss value. In order to check the predictive ability of the different
models, we report the AE and MAE estimated
from section 3.3.5.\\

Table 3 reports the AE between actual portfolio loss and expected
portfolio loss constructed by 31 and 62 observations in 2008 and 2009,
respectively. In a comparison with
four models, one can observe that the estimate of expected portfolio loss in
the RRFL model is highest and closest to the corresponding actual one, which means the expected portfolio losses
may be underestimated by the other three models. Especially, a modelling recovery rate in a stochastic fashion indeed contributes to credit loss estimation.

We compare the four competing models of each obligor and choose
the best model which achieves the minimum AE and MAE. It can be seen that
including the conditional factor loading (RFL model) instead of the Spearman
correlation (FC model) does not significantly improve the estimations in 2008 and 2009. As can be seen in Table 4, we find that introducing
the state-dependent recovery rate (RR model) leads to a promising improvement
over the standard model (FC model). We interpret this as saying that the setting of stochastic recovery rate seems necessary, which brings a remarkable improvement on default prediction. This result is consistent with {\color{blue}\citet{altman2005link}} and {\color{blue}\citet{ferreira2007corporate}}. Compared with the RR model, the RRFL model
includes conditional factor loading in default probabilities and
a state-dependent recovery rates function which produces much more
modest improvements.\\

We propose two specifications on factor loading, and recovery rates across four models. If we assume that default probabilities are
the function of two-state correlation constructers, but recovery
rates do not, the specification is only identified concentrated on factor loading. In this case, the recovery rates do not
contain information about the state of business cycle. Conversely, if
we assume that recovery rates vary, but factor loading is fixed, then the refinement is only through the
variation in the recovery rate. Since the RRFL model with both specifications is superior to the
other three competing models, there is no redundant specification in this study. In this regard, we extend the models proposed by prior literatures ({\color{blue}\citealp{van2007factor}}; {\color{blue}\citealp{rosen2010risk}}; {\color{blue}\citealp{kalemanova2007normal}}; {\color{blue}\citealp{chen2014mixed}}; {\color{blue}\citealp{amraoui2008optimal}};  
{\color{blue}\citealp{elouerkhaoui2009base}}; {\color{blue}\citealp{amraoui2012pricing}}) which leads more accurate default prediction in one year.  
\begin{table}[!ht]
    \begin{center}
        \begin{tabular}{lrrrrr}
            \hline\hline
            &FC & RFL & RR & RRFL\\
            \hline
            2008\\
            \hline
            Actual portfolio loss & 2035.02 &  2035.02 &  2035.02 &  2035.02 \\
            Expected portfolio loss & 509.60& 527.06& 687.01& 690.86 \\
            AE & 1525.42 & 1507.96 & 1348.01 & 1344.16 \\
            MAE & 47.12 & 47.67 & 42.13 & 42.01 \\
            Expected portfolio loss/Actual portfolio loss & 25.04\%&25.90\%&33.76\%&33.95\%&  \\
            \hline
            2009\\
            \hline
             Actual portfolio loss & 4073.80 &  4073.80 &  4073.80 &  4073.80 \\
             Expected portfolio loss & 1203.56& 1212.38& 1769.14& 1788.05 \\
             AE & 2870.24 & 2861.42 & 2304.67 & 2285.75 \\
             MAE & 43.49 & 43.35 & 34.92 & 34.63\\
             Expected portfolio loss/Actual portfolio loss & 29.54\%&29.76\%&43.43\%&43.89\%&  \\
            \hline \hline
        \end{tabular}
    \end{center}
    \caption{{\bf The mean of actual portfolio loss, expected portfolio loss  and AE, MAE (in million)}\\This table reports the
    	AE and MAE by comparing the four models: (1) The FC model: the
    	standard one-factor Gaussian copula model with the constant
    	recovery rate. (2) The RFL model: the
    	one-factor Gaussian copula model with the factor loadings tied to
    	the state of common factor and the recoveries being assumed to be constant. (3) The RR model:
    	standard one-factor Gaussian copula model but the recoveries being
    	related to the state of the macroeconomic state. and (4)The RRFL
    	model: a conditional factor
    	loading specification together with a state-dependent recovery rate. This table also presents the difference between actual portfolio loss and expected portfolio loss as referred to AE, and divided by 31 and 62 observations in 2008 and 2009, respectively, as MAE. The percentage represents expected portfolio loss divided by the actual portfolio loss.\\
    }\label{table4}
\end{table}

\subsection{Basel III: Relative contribution}

Since Basel III is proposed to control systematic risk (one of systemic risk measures) to
achieve the goal of overall financial stability, systematic risk
has been considered one of the main causes of the 2007-2009 crisis.
In this section, we highlight the role of systematic risk and its
impact to fit the
goals of Basel III. The aim of relative contribution analysis is to
investigate the proportional contribution from systematic risk in
comparison to that from the idiosyncratic component. By measuring
the systematic risk, $\alpha_i^{S}Z$, and idiosyncratic risk,
$\sqrt{1-(\alpha^{S}_i)^2}\varepsilon_i$,
$\mbox{S}\in\{\mbox{H,Q}\}$ from formula (6,7), we depict a
scatter plot for simulated systematic risk (horizontal axis) and
idiosyncratic risk (vertical axis) in Figure 5. As can be seen in the 2D plot in 2008, the $45^{\circ}$ line represents the proportion of systematic risk is equal to that of idiosyncratic risk. If scattered
points are located in the \textquoteleft A, B, C, D\textquoteright\, zones, the contribution of
systematic risk on default risk is greater than idiosyncratic risk. On the other hand, if scattered points are settled in the \textquoteleft a, b, c, d\textquoteright\, areas, the contribution of
systematic component is less than idiosyncratic risk. For example, the effect of systematic risk on default risk will become larger when point \textquoteleft Y\textquoteright\, moves to point \textquoteleft X\textquoteright. Most literature focus on 
either systematic ({\color{blue} \citealp{huang2009framework}};
{\color{blue} \citealp{acharya2010measuring}}) or firm-specific
components ({\color{blue} \citealp{goyal2003idiosyncratic}};
{\color{blue}
\citealp{ferreira2007corporate}}), and a limit number of studies compare the influence of both of them. \\

 By simulating $Z \sim \textsf{N}(-0.03, 3.05)$, each simulated Z random variable can therefore be mapped into a specific conditional probability of being hectic state in Eq. (12). We depict the scatters in three groups here. The first group (marked as \textquoteleft +\textquoteright\, in green) only includes the simulated Z r.v. with projecting conditional probabilities above $75\%$-quartile, and indicates that they are generated in distress.  The second group (marked as \textquoteleft *\textquoteright\, in blue) includes the Z r.v. with projecting conditional probabilities below $25\%$-quartile to indicate that they are generated in a bullish atmosphere. The third group (marked as \textquoteleft x\textquoteright\, in red) collects the rest. With regards to the tranquil
scenarios (\textquoteleft blue \textquoteright\, points) in 2008, most observations were located in the area where
the relative contribution of idiosyncratic risk is larger than the
economy-wide component, that the credit risk was mainly
driven by the idiosyncratic component before the subprime crisis as reported in the {\color{blue}\citet{rodriguez2013systemic}} article. In their article, they
find that idiosyncratic components are larger than systematic risk
before the subprime crisis, extracted from the CDX-IG-5y by using
high-frequent measures. At the beginning of the financial crisis,
systematic risk skyrocketed. Intuitively, the systematic risk increases sharply due to the larger factor loadings when the market is in hectic scenarios. Our result shows systematic risk is
higher than the idiosyncratic component in the hectic scenarios (\textquoteleft green \textquoteright\,points) in 2008; in
the quiet scenarios, firm-specific factors, however, are mostly important at some
points, as proposed by
{\color{blue}\citet{rodriguez2013systemic}}. Similarly, it has
been shown that the relative contribution of the systematic component
explains a higher proportion of obligor asset value in 2009.\\

More visibly, the 3D plot identifies the relationship among the
level of average $U_{i}|_{\mbox{S}}$ referred to as the mean of firms' value,
systematic and idiosyncratic component. Each observation in Figure
4 reflects its mean of $U_{i}|_{\mbox{S}}$ $i=1,\ldots,N$ in each simulating
day in 2008 and 2009, respectively. As can be seen in Figure 5, the 
points in the hectic period marked as green circles indicates a
negative shock from systematic risk which lowers the average asset
value of obligors; specifically, the majority of observations show
a negative impact of systematic shock which accounts for a larger
proportion on the firms' values substantially. Note that it is easy to drive the default event since it lowers
the firms' value significantly. On the
other hand, the points in quiet days marked as blue circles
indicate a positive shock from the systematic component. However,
the negative shock from firm-specific factors
may compromise the benefit from economy-wide components that lowers the
level of average $U_{i}|_{\mbox{S}}$ at some points.\\

Our model emphasises the importance of systematic risk which
explains most obligors default behaviour particular in hectic
periods, which is one of the important measures of Basel III
({\color{blue} \citealp{schwerter2011basel}}; {\color{blue} \citealp{uhde2010securitization}}; {\color{blue} \citealp{tarashev2010attributing}}). To be specific, we measure and
demonstrate the contribution of overall systematic risk to each
asset, and identify the impact direction from systematic and idiosyncratic
risk. Moreover, it can be applied to a variety of systematic risk
measures. In this sense, portfolio managers should be aware of the
systematic risk which influences the value of portfolios
substantially. We propose that the regulatory tool of Basel III
could be estimated according to such
contributions. A related question is how these measures can aid policymakers. The
measures in this paper can be used as a tool to prevent systematic
crisis. Our model can be used as an early warning system that will
alert the regulators when an individual bank is in trouble and to intervene before the crisis happens.\\

\begin{figure}[!ht]
    \begin{subfigure}{0.5\textwidth}
       \begin{subfigure}{\textwidth}
          \renewcommand\thesubfigure{\alph{subfigure}}
          \centering
          \includegraphics[scale=0.45]{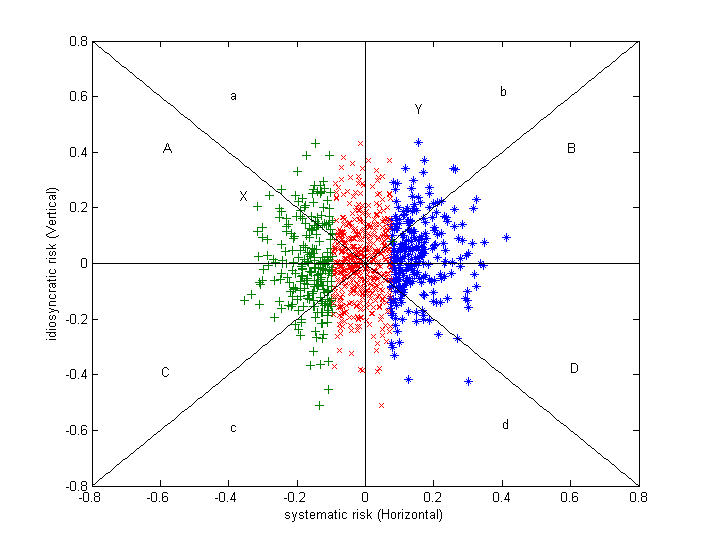}
       \end{subfigure}
       \begin{subfigure}{\textwidth}
          \renewcommand\thesubfigure{\alph{subfigure}}
          \centering
          \includegraphics[scale=0.45]{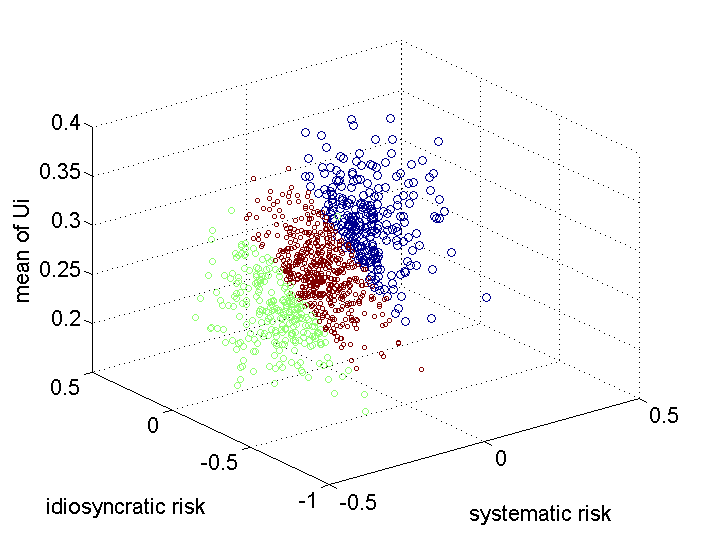}
       \end{subfigure}
        \caption{2008}
    \end{subfigure}
    \begin{subfigure}{0.5\textwidth}
       \begin{subfigure}{\textwidth}
          \renewcommand\thesubfigure{\alph{subfigure}}
          \centering
          \includegraphics[scale=0.45]{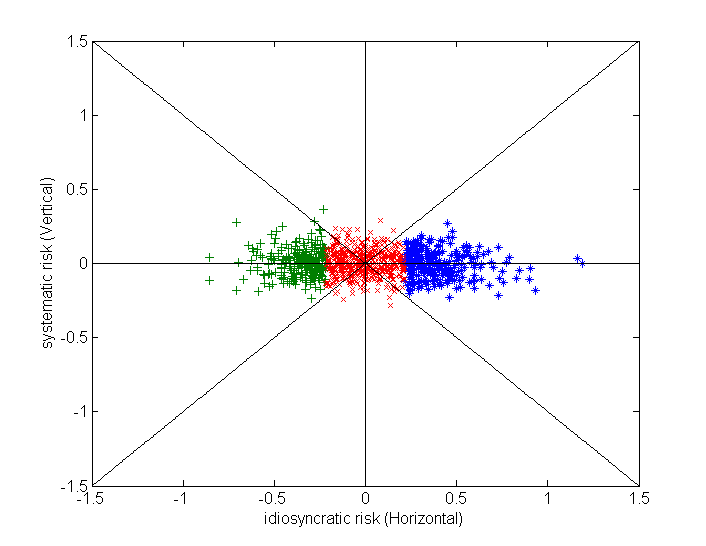}
       \end{subfigure}
       \begin{subfigure}{\textwidth}
          \renewcommand\thesubfigure{\alph{subfigure}}
          \centering
          \includegraphics[scale=0.45]{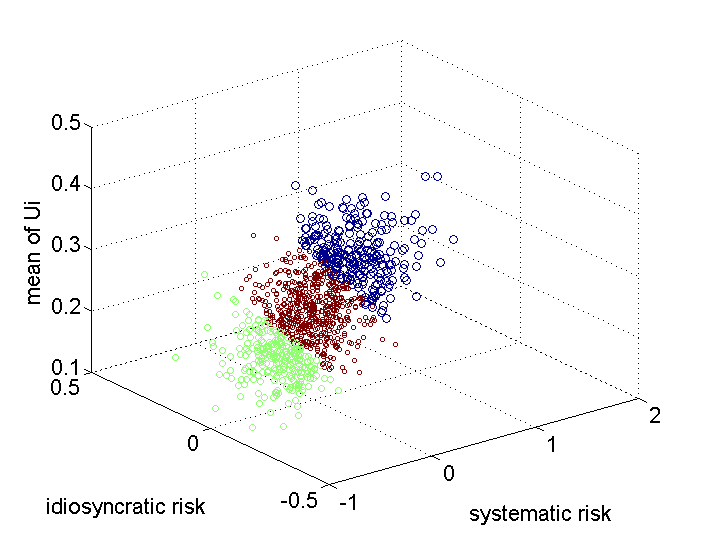}
       \end{subfigure}
       \caption{2009}
    \end{subfigure}
  \caption{{\bf The 2D and 3D scatters plot of relative
  contribution}\\
           By simulating $Z \sim \textsf{N}(-0.03, 3.05)$, the 2D graphic illustrates the relationship between the mean of systematic risk, $\alpha_iZ$, and idiosyncratic risk, $\sqrt{1-\alpha_i^2}\varepsilon_i$. Each simulated Z random variable can therefore be mapped into a specific conditional probability of being hectic state in Eq. (12). We depict the scatters in three groups here. The first group (marked as \textquoteleft +\textquoteright\, in green) only includes the simulated Z r.v. with projecting conditional probabilities above $75\%$-quartile, and indicates that they are generated in distress.  The second group (marked as \textquoteleft *\textquoteright\, in blue) includes the Z r.v. with projecting conditional probabilities below $25\%$-quartile to indicate that they are generated in a bullish atmosphere. The third group (marked as \textquoteleft x\textquoteright\, in red) collects the rest. In 3D plot, observations
           in hectic periods are marked as green circles. In quiet
           days are marked as blue circles, otherwise as red
           circles.
          } \label{figure5}
\end{figure}

\subsection{Robustness test}

 Since Table 3 reports that the expected portfolio loss is far away from the actual portfolio loss, we gauge that using bond credit rates as a measure of hazard rate has the disadvantage that they are released annually by Moody's report. In this section, we use credit default swap (CDS) spread data as an alternative  market-based measure of the company's credit risk. A CDS spread is a financial swap agreement that the seller of CDS will compensate the buyer in the event of a loan default. Basically, the variation of CDS spread reflects the dynamic of risk condition or hazard rate implicitly. The larger the CDS spread is, the riskier the debtor is. Therefore, the hazard rate, $\bar{\kappa}$, for a company can be estimated by,\\

\begin{equation}
\bar{\kappa}=\frac{s}{1-R}
\end{equation}\\

\noindent where $s$ is CDS spread. We consider the latest one-year CDS quotes of obligors before the default year provided from Datastream. We also use a credit spread which is the yield on a annual par yield bond issued by the obligors over one-year LIBOR (London Interbank Offered Rate) if the obligor doesn't have CDS data. Theoretically, the CDS spread is very close to the credit spread ({\color{blue} \citealp{hull2000valuing}}; {\color{blue} \citealp{hull2004merton}}). By plugging in the recovery rate, $R$, obtained from Moody's report, we compute the
average default intensity, $\bar{\kappa}$, per year conditional on no earlier
default instead of $P_i$. Compared with $P_i$ from Moody's annual report, a CDS spread with active trading activity reflects
market assessments of default risk in a timely fashion. In this regard, the proposed models with an incorporation of the hazard rate implied in CDS spreads may produce a better prediction.\\

According to Table 4, the models with a hazard rate implied in a CDS
spread seem to perform better than those with a hazard rate from historical
bond credit rates. By comparing Tables 3 and 4, generally, a CDS spread as the hazard rate measure reflects information more timely than the bond credit rate does. As can be seen in Table 4, the RRFL model outperforms in robustness test. In both Tables, the RRFL model consistently outperforms, which produces the expected portfolio loss most closely to the actual portfolio loss.\\ 
\begin{table}[!ht]
    \begin{center}
        \begin{tabular}{lrrrrr}
            \hline\hline
            &FC & RFL & RR & RRFL\\
            \hline
            2008\\
            \hline
            Actual portfolio loss & 1401.31 &  1401.31 &  1401.31 &  1401.31 \\
            Expected portfolio loss & 560.50& 533.82& 589.54& 591.40 \\
            AE & 840.81 & 867.49 & 811.77 & 809.91 \\
            MAE & 35.03 & 36.15 & 33.82 & 33.75 \\
            Expected portfolio loss/Actual portfolio loss & 40.00\%&38.09\%&42.07\%&42.20\%&  \\
            \hline
            2009\\
            \hline
            Actual portfolio loss & 2707.30 &  2707.30 & 2707.30 &  2707.30 \\
            Expected portfolio loss & 1457.07& 1462.18& 1677.89& 1683.97 \\
            AE & 1250.23 & 1245.12 & 1029.42 & 1023.33 \\
            MAE & 29.77 & 29.65 & 24.51 & 24.37\\
            Expected portfolio loss/Actual portfolio loss & 53.82\%&54.01\%&61.98\%&62.20\%&  \\
            \hline \hline
        \end{tabular}
    \end{center}
    \caption{{\bf The actual portfolio loss, expected portfolio loss, AE, and MAE (in million) for robustness}\\This table reports the value of AE and MAE of four models by using market-based method during 2008 and 2009. This table also shows the actual portfolio loss and expected portfolio loss of 24 and 42 observations in 2008 and 2009. The percentage represents expected portfolio loss divided by the actual portfolio loss.\\
    }\label{table5}
\end{table}


\section{Conclusion}

This paper proposes a refined factor copula model for credit
risk prediction. On the basis of our estimated model, we find that
systematic risk plays a critical role in governing default
rates and recovery rates simultaneously. Our simulation results
show that recoveries vary with the returns of the S\&P 500 and the
impact of systematic factors on the recovery rate is asymmetric by characterising a higher factor loading in hectic periods than in tranquil ones. Among the various factor copula models developed in the past and current literature as the competing models, the one with conditional random factor loading and a state-dependent recovery rate turns out to be the most superior. In other words, our refined model contributes to literature that have been mapped to 3 groups of competing models (the FC, RFL, and RR models)\\

As a response to Basel III, we measure and demonstrate the
contribution of overall systematic risk to each firm's value and
identify the relative role of the systematic and idiosyncratic risk.
Moreover, it can be applied to a variety of systematic risk
measures, and aids regulators in preventing a systematic crisis. In
addition, by investigating the effect of state-dependent recovery
rates on the loss function, we suggest that banks should apply this issue on capital requirement to make sure of its sufficiency.\\

In further research, we plan to go beyond this study in several ways. For instance, other copula functions can be modelled to capture various dependence structures. Secondly, the marginal distribution can be considered in a more general way to capture a fat-tail feature. We will leave these issues for future studies.\\

\section*{Acknowledgements} 
This research was financially
supported by the Deutsche Forschungsgemeinschaft (DFG) via SFB 649
"\"{O}konomisches Risiko" and IRTG 1792 "High-Dimensional
Non-Stationary Times Series" is gratefully acknowledged.


\begin{appendices}

\section{Conditional Factor Loading}
\setcounter{equation}{0}
\renewcommand{\theequation}{A.\arabic{equation}}

We assume the two asset returns $Z$ (USD S\&P 500), $U_i$ (firm
stock price) to have a mixture of bivariate normal distribution:\\
$(Z,U_i)\sim$
\begin{equation}\hspace{-0.50cm} \left\{
\begin{array}{cc}
N\left\{
\begin{array}{cc}
\left[
\begin{array}{c}
\mu^{Q}_Z \\
\mu^{Q}_{i}%
\end{array}%
\right] , & \left[
\begin{array}{cc}
(\sigma^{Q}_Z)^{2} & (\sigma^{Q}_Z)^{{}}(\alpha^{Q})(\sigma^{Q}_{i})^{{}} \\
(\sigma^{Q}_Z)^{{}}(\alpha^{Q})(\sigma^{Q}_{i})^{{}} & (\sigma^{Q}_{i})^{2}%
\end{array}%
\right]
\end{array}%
\right\}  & \operatorname{P}(\mbox{S=Q})=1-\omega  \\
N\left\{
\begin{array}{cc}
\left[
\begin{array}{c}
\mu^{H}_Z \\
\mu^{H}_{i}%
\end{array}%
\right] , & \left[
\begin{array}{cc}
(\sigma^{H}_Z)^{2} & (\sigma^{H}_Z)^{{}}(\alpha^{H})(\sigma^{H}_{i})^{{}} \\
(\sigma^{H}_Z)^{{}}(\alpha^{H})(\sigma^{H}_{i})^{{}} & (\sigma^{H}_{i})^{2}%
\end{array}%
\right]
\end{array}%
\right\}  & \operatorname{P}(\mbox{S=H})=\omega
\end{array}%
\right.
\end{equation}\\

\noindent where volatility in hectic periods is higher than in a
quiet periods, $(\sigma
^{H}_i)^{2}>(\sigma^{Q}_i)^{2}$. $\alpha^{Q}$ and $\alpha^{H}$ are the correlation coefficient between each obligor and the S\&P 500 in quiet and hectic period proposed by {\color{blue}\citet{kim2000stress}}, respectively.\\
We estimate the unknown parameters $\omega$, $\mu^{Q}_Z$, $\sigma
^{Q}_Z$, $\mu^{H}_Z$, $\sigma^{H}_Z$ from the marginal
distribution of $Z$:\\

\begin{equation} \left\{
\begin{array}{cc}
N\left[
\begin{array}{cc}
\mu^{Q}_Z, & (\sigma
^{Q}_Z)^{2}%
\end{array}%
\right]  & \operatorname{P}(\mbox{S=Q})=1-\omega  \\
N\left[
\begin{array}{cc}
\mu^{H}_Z, & (\sigma^{H}_Z)^{2}%
\end{array}%
\right]  & \operatorname{P}(\mbox{S=H})=\omega
\end{array}%
\right.
\end{equation}
\end{appendices}
\bibliographystyle{model2-names}
\bibliography{barrier}

\begin{thebibliography}{45}
\expandafter\ifx\csname natexlab\endcsname\relax\def\natexlab#1{#1}\fi
\providecommand{\url}[1]{\texttt{#1}}
\providecommand{\href}[2]{#2}
\providecommand{\path}[1]{#1}
\providecommand{\DOIprefix}{doi:}
\providecommand{\ArXivprefix}{arXiv:}
\providecommand{\URLprefix}{URL: }
\providecommand{\Pubmedprefix}{pmid:}
\providecommand{\doi}[1]{\href{http://dx.doi.org/#1}{\path{#1}}}
\providecommand{\Pubmed}[1]{\href{pmid:#1}{\path{#1}}}
\providecommand{\bibinfo}[2]{#2}
\ifx\xfnm\relax \def\xfnm[#1]{\unskip,\space#1}\fi
\bibitem[{Acharya et~al.(2010)Acharya, Pedersen, Philippon and
  Richardson}]{acharya2010measuring}
\bibinfo{author}{Acharya, V.V.}, \bibinfo{author}{Pedersen, L.H.},
  \bibinfo{author}{Philippon, T.}, \bibinfo{author}{Richardson, M.P.},
  \bibinfo{year}{2010}.
\newblock \bibinfo{title}{Measuring systemic risk} .
\bibitem[{Altman et~al.(2005)Altman, Brady, Resti and Sironi}]{altman2005link}
\bibinfo{author}{Altman, E.I.}, \bibinfo{author}{Brady, B.},
  \bibinfo{author}{Resti, A.}, \bibinfo{author}{Sironi, A.},
  \bibinfo{year}{2005}.
\newblock \bibinfo{title}{The link between default and recovery rates: Theory,
  empirical evidence, and implications*}.
\newblock \bibinfo{journal}{The Journal of Business} \bibinfo{volume}{78},
  \bibinfo{pages}{2203--2228}.
\bibitem[{Amraoui et~al.(2012)Amraoui, Cousot, Hitier and
  Laurent}]{amraoui2012pricing}
\bibinfo{author}{Amraoui, S.}, \bibinfo{author}{Cousot, L.},
  \bibinfo{author}{Hitier, S.}, \bibinfo{author}{Laurent, J.P.},
  \bibinfo{year}{2012}.
\newblock \bibinfo{title}{Pricing {CDOs} with state-dependent stochastic
  recovery rates}.
\newblock \bibinfo{journal}{Quantitative Finance} \bibinfo{volume}{12},
  \bibinfo{pages}{1219--1240}.
\bibitem[{Amraoui and Hitier(2008)}]{amraoui2008optimal}
\bibinfo{author}{Amraoui, S.}, \bibinfo{author}{Hitier, S.},
  \bibinfo{year}{2008}.
\newblock \bibinfo{title}{Optimal stochastic recovery for base correlation}.
\newblock \bibinfo{journal}{BNP Paribas} .
\bibitem[{Andersen and Sidenius(2004)}]{andersen2004extensions}
\bibinfo{author}{Andersen, L.B.}, \bibinfo{author}{Sidenius, J.},
  \bibinfo{year}{2004}.
\newblock \bibinfo{title}{Extensions to the gaussian copula: Random recovery
  and random factor loadings}.
\newblock \bibinfo{journal}{Journal of Credit Risk} \bibinfo{volume}{1},
  \bibinfo{pages}{29--70}.
\bibitem[{Ang and Bekaert(2002a)}]{ang2002international}
\bibinfo{author}{Ang, A.}, \bibinfo{author}{Bekaert, G.},
  \bibinfo{year}{2002a}.
\newblock \bibinfo{title}{International asset allocation with regime shifts}.
\newblock \bibinfo{journal}{Review of Financial studies} \bibinfo{volume}{15},
  \bibinfo{pages}{1137--1187}.
\bibitem[{Ang and Chen(2002b)}]{ang2002asymmetric}
\bibinfo{author}{Ang, A.}, \bibinfo{author}{Chen, J.}, \bibinfo{year}{2002b}.
\newblock \bibinfo{title}{Asymmetric correlations of equity portfolios}.
\newblock \bibinfo{journal}{Journal of Financial Economics}
  \bibinfo{volume}{63}, \bibinfo{pages}{443--494}.
\bibitem[{Bonti et~al.(2006)Bonti, Kalkbrener, Lotz and
  Stahl}]{bonti2006credit}
\bibinfo{author}{Bonti, G.}, \bibinfo{author}{Kalkbrener, M.},
  \bibinfo{author}{Lotz, C.}, \bibinfo{author}{Stahl, G.},
  \bibinfo{year}{2006}.
\newblock \bibinfo{title}{Credit risk concentrations under stress}.
\newblock \bibinfo{journal}{Journal of Credit Risk} \bibinfo{volume}{2},
  \bibinfo{pages}{115--136}.
\bibitem[{Bruche and Gonzalez-Aguado(2010)}]{bruche2010recovery}
\bibinfo{author}{Bruche, M.}, \bibinfo{author}{Gonzalez-Aguado, C.},
  \bibinfo{year}{2010}.
\newblock \bibinfo{title}{Recovery rates, default probabilities, and the credit
  cycle}.
\newblock \bibinfo{journal}{Journal of Banking \& Finance}
  \bibinfo{volume}{34}, \bibinfo{pages}{754--764}.
\bibitem[{Carty et~al.(1998)Carty, Hamilton, Keenan, Moss, Mulvaney, Marshella
  and Subhas}]{carty1998bankrupt}
\bibinfo{author}{Carty, L.V.}, \bibinfo{author}{Hamilton, D.T.},
  \bibinfo{author}{Keenan, S.C.}, \bibinfo{author}{Moss, A.},
  \bibinfo{author}{Mulvaney, M.}, \bibinfo{author}{Marshella, T.},
  \bibinfo{author}{Subhas, M.}, \bibinfo{year}{1998}.
\newblock \bibinfo{title}{Bankrupt bank loan recoveries}.
\newblock \bibinfo{journal}{Moody’s investors service} \bibinfo{volume}{15},
  \bibinfo{pages}{79}.
\bibitem[{Chen(2010)}]{chen2010macroeconomic}
\bibinfo{author}{Chen, H.}, \bibinfo{year}{2010}.
\newblock \bibinfo{title}{Macroeconomic conditions and the puzzles of credit
  spreads and capital structure}.
\newblock \bibinfo{journal}{The Journal of Finance} \bibinfo{volume}{65},
  \bibinfo{pages}{2171--2212}.
\bibitem[{Chen et~al.(2014)Chen, Liu and Li}]{chen2014mixed}
\bibinfo{author}{Chen, J.}, \bibinfo{author}{Liu, Z.}, \bibinfo{author}{Li,
  S.}, \bibinfo{year}{2014}.
\newblock \bibinfo{title}{Mixed copula model with stochastic correlation for
  cdo pricing}.
\newblock \bibinfo{journal}{Economic Modelling} \bibinfo{volume}{40},
  \bibinfo{pages}{167--174}.
\bibitem[{Choro{\'s}-Tomczyk et~al.(2013)Choro{\'s}-Tomczyk, H{\"a}rdle and
  Okhrin}]{choros2013valuation}
\bibinfo{author}{Choro{\'s}-Tomczyk, B.}, \bibinfo{author}{H{\"a}rdle, W.K.},
  \bibinfo{author}{Okhrin, O.}, \bibinfo{year}{2013}.
\newblock \bibinfo{title}{Valuation of collateralized debt obligations with
  hierarchical archimedean copulae}.
\newblock \bibinfo{journal}{Journal of Empirical Finance} \bibinfo{volume}{24},
  \bibinfo{pages}{42--62}.
\bibitem[{Choro{\'s}-Tomczyk et~al.(2014)Choro{\'s}-Tomczyk, H{\"a}rdle and
  Overbeck}]{choros2014copula}
\bibinfo{author}{Choro{\'s}-Tomczyk, B.}, \bibinfo{author}{H{\"a}rdle, W.K.},
  \bibinfo{author}{Overbeck, L.}, \bibinfo{year}{2014}.
\newblock \bibinfo{title}{Copula dynamics in {CDOs}}.
\newblock \bibinfo{journal}{Quantitative Finance} \bibinfo{volume}{14},
  \bibinfo{pages}{1573--1585}.
\bibitem[{Committee et~al.(2009)}]{basel2009strengthening}
\bibinfo{author}{Committee, B.}, et~al., \bibinfo{year}{2009}.
\newblock \bibinfo{title}{Strengthening the resilience of the banking sector}.
\newblock \bibinfo{journal}{Basel Committee} .
\bibitem[{Crouhy et~al.(2000)Crouhy, Galai and Mark}]{crouhy2000comparative}
\bibinfo{author}{Crouhy, M.}, \bibinfo{author}{Galai, D.},
  \bibinfo{author}{Mark, R.}, \bibinfo{year}{2000}.
\newblock \bibinfo{title}{A comparative analysis of current credit risk
  models}.
\newblock \bibinfo{journal}{Journal of Banking \& Finance}
  \bibinfo{volume}{24}, \bibinfo{pages}{59--117}.
\bibitem[{Das and Hanouna(2009)}]{das2009hedging}
\bibinfo{author}{Das, S.R.}, \bibinfo{author}{Hanouna, P.},
  \bibinfo{year}{2009}.
\newblock \bibinfo{title}{Hedging credit: Equity liquidity matters}.
\newblock \bibinfo{journal}{Journal of Financial Intermediation}
  \bibinfo{volume}{18}, \bibinfo{pages}{112--123}.
\bibitem[{Drehmann and Tarashev(2013)}]{drehmann2013measuring}
\bibinfo{author}{Drehmann, M.}, \bibinfo{author}{Tarashev, N.},
  \bibinfo{year}{2013}.
\newblock \bibinfo{title}{Measuring the systemic importance of interconnected
  banks}.
\newblock \bibinfo{journal}{Journal of Financial Intermediation}
  \bibinfo{volume}{22}, \bibinfo{pages}{586--607}.
\bibitem[{Elouerkhaoui(2009)}]{elouerkhaoui2009base}
\bibinfo{author}{Elouerkhaoui, Y.}, \bibinfo{year}{2009}.
\newblock \bibinfo{title}{Base correlation calibration with a stochastic
  recovery model}.
\newblock \bibinfo{type}{Technical Report}. working paper, Citigroup Global
  Markets.
\bibitem[{Ferreira and Laux(2007)}]{ferreira2007corporate}
\bibinfo{author}{Ferreira, M.A.}, \bibinfo{author}{Laux, P.A.},
  \bibinfo{year}{2007}.
\newblock \bibinfo{title}{Corporate governance, idiosyncratic risk, and
  information flow}.
\newblock \bibinfo{journal}{The Journal of Finance} \bibinfo{volume}{62},
  \bibinfo{pages}{951--989}.
\bibitem[{Franke et~al.(2015)Franke, H{\"a}rdle and
  Hafner}]{franke2015statistics}
\bibinfo{author}{Franke, J.}, \bibinfo{author}{H{\"a}rdle, W.K.},
  \bibinfo{author}{Hafner, C.M.}, \bibinfo{year}{2015}.
\newblock \bibinfo{title}{Statistics of financial markets: an introduction}.
\newblock \bibinfo{publisher}{Springer Science \& Business Media}.
\bibitem[{Frey and McNeil(2003)}]{frey2003dependent}
\bibinfo{author}{Frey, R.}, \bibinfo{author}{McNeil, A.J.},
  \bibinfo{year}{2003}.
\newblock \bibinfo{title}{Dependent defaults in models of portfolio credit
  risk}.
\newblock \bibinfo{journal}{Journal of Risk} \bibinfo{volume}{6},
  \bibinfo{pages}{59--92}.
\bibitem[{Goyal and Santa-Clara(2003)}]{goyal2003idiosyncratic}
\bibinfo{author}{Goyal, A.}, \bibinfo{author}{Santa-Clara, P.},
  \bibinfo{year}{2003}.
\newblock \bibinfo{title}{Idiosyncratic risk matters!}
\newblock \bibinfo{journal}{The Journal of Finance} \bibinfo{volume}{58},
  \bibinfo{pages}{975--1008}.
\bibitem[{Huang et~al.(2009)Huang, Zhou and Zhu}]{huang2009framework}
\bibinfo{author}{Huang, X.}, \bibinfo{author}{Zhou, H.}, \bibinfo{author}{Zhu,
  H.}, \bibinfo{year}{2009}.
\newblock \bibinfo{title}{A framework for assessing the systemic risk of major
  financial institutions}.
\newblock \bibinfo{journal}{Journal of Banking \& Finance}
  \bibinfo{volume}{33}, \bibinfo{pages}{2036--2049}.
\bibitem[{Hull et~al.(2004)Hull, Nelken and White}]{hull2004merton}
\bibinfo{author}{Hull, J.}, \bibinfo{author}{Nelken, I.},
  \bibinfo{author}{White, A.}, \bibinfo{year}{2004}.
\newblock \bibinfo{title}{Merton’s model, credit risk, and volatility skews}.
\newblock \bibinfo{journal}{Journal of Credit Risk Volume} \bibinfo{volume}{1},
  \bibinfo{pages}{03--27}.
\bibitem[{Hull(2006)}]{hull2006options}
\bibinfo{author}{Hull, J.C.}, \bibinfo{year}{2006}.
\newblock \bibinfo{title}{Options, futures, and other derivatives}.
\newblock \bibinfo{publisher}{Pearson Education India}.
\bibitem[{Hull and White(2000)}]{hull2000valuing}
\bibinfo{author}{Hull, J.C.}, \bibinfo{author}{White, A.},
  \bibinfo{year}{2000}.
\newblock \bibinfo{title}{Valuing credit default swaps i: No counterparty
  default risk} .
\bibitem[{Hull and White(2004)}]{hull2004valuation}
\bibinfo{author}{Hull, J.C.}, \bibinfo{author}{White, A.D.},
  \bibinfo{year}{2004}.
\newblock \bibinfo{title}{Valuation of a {CDO} and an $n$-th to default {CDS}
  without monte carlo simulation}.
\newblock \bibinfo{journal}{The Journal of Derivatives} \bibinfo{volume}{12},
  \bibinfo{pages}{8--23}.
\bibitem[{Jarrow et~al.(1997)Jarrow, Lando and Turnbull}]{jarrow1997markov}
\bibinfo{author}{Jarrow, R.A.}, \bibinfo{author}{Lando, D.},
  \bibinfo{author}{Turnbull, S.M.}, \bibinfo{year}{1997}.
\newblock \bibinfo{title}{A markov model for the term structure of credit risk
  spreads}.
\newblock \bibinfo{journal}{Review of financial studies} \bibinfo{volume}{10},
  \bibinfo{pages}{481--523}.
\bibitem[{Kalemanova et~al.(2007)Kalemanova, Schmid and
  Werner}]{kalemanova2007normal}
\bibinfo{author}{Kalemanova, A.}, \bibinfo{author}{Schmid, B.},
  \bibinfo{author}{Werner, R.}, \bibinfo{year}{2007}.
\newblock \bibinfo{title}{The normal inverse gaussian distribution for
  synthetic {CDO} pricing}.
\newblock \bibinfo{journal}{The Journal of Derivatives} \bibinfo{volume}{14},
  \bibinfo{pages}{80--94}.
\bibitem[{Khieu et~al.(2012)Khieu, Mullineaux and Yi}]{khieu2012determinants}
\bibinfo{author}{Khieu, H.D.}, \bibinfo{author}{Mullineaux, D.J.},
  \bibinfo{author}{Yi, H.C.}, \bibinfo{year}{2012}.
\newblock \bibinfo{title}{The determinants of bank loan recovery rates}.
\newblock \bibinfo{journal}{Journal of Banking \& Finance}
  \bibinfo{volume}{36}, \bibinfo{pages}{923--933}.
\bibitem[{Kim and Finger(2000)}]{kim2000stress}
\bibinfo{author}{Kim, J.}, \bibinfo{author}{Finger, C.C.},
  \bibinfo{year}{2000}.
\newblock \bibinfo{title}{A stress test to incorporate correlation breakdown}.
\newblock \bibinfo{journal}{Journal of Risk} \bibinfo{volume}{2},
  \bibinfo{pages}{5--20}.
\bibitem[{Longin and Solnik(2001)}]{longin2001extreme}
\bibinfo{author}{Longin, F.}, \bibinfo{author}{Solnik, B.},
  \bibinfo{year}{2001}.
\newblock \bibinfo{title}{Extreme correlation of international equity markets}.
\newblock \bibinfo{journal}{The Journal of Finance} \bibinfo{volume}{56},
  \bibinfo{pages}{649--676}.
\bibitem[{Merton(1974)}]{merton1974pricing}
\bibinfo{author}{Merton, R.C.}, \bibinfo{year}{1974}.
\newblock \bibinfo{title}{On the pricing of corporate debt: The risk structure
  of interest rates*}.
\newblock \bibinfo{journal}{The Journal of Finance} \bibinfo{volume}{29},
  \bibinfo{pages}{449--470}.
\bibitem[{Pagratis and Stringa(2009)}]{pagratis2009modeling}
\bibinfo{author}{Pagratis, S.}, \bibinfo{author}{Stringa, M.},
  \bibinfo{year}{2009}.
\newblock \bibinfo{title}{Modeling bank senior unsecured ratings: a reasoned
  structured approach to bank credit assessment}.
\newblock \bibinfo{journal}{International Journal of Central Banking}
  \bibinfo{volume}{5}, \bibinfo{pages}{1--39}.
\bibitem[{Pan and Singleton(2008)}]{pan2008default}
\bibinfo{author}{Pan, J.}, \bibinfo{author}{Singleton, K.J.},
  \bibinfo{year}{2008}.
\newblock \bibinfo{title}{Default and recovery implicit in the term structure
  of sovereign {CDS} spreads}.
\newblock \bibinfo{journal}{The Journal of Finance} \bibinfo{volume}{63},
  \bibinfo{pages}{2345--2384}.
\bibitem[{Patton(2004)}]{patton2004out}
\bibinfo{author}{Patton, A.J.}, \bibinfo{year}{2004}.
\newblock \bibinfo{title}{On the out-of-sample importance of skewness and
  asymmetric dependence for asset allocation}.
\newblock \bibinfo{journal}{Journal of Financial Econometrics}
  \bibinfo{volume}{2}, \bibinfo{pages}{130--168}.
\bibitem[{Pykhtin and Dev(2002)}]{pykhtin2002credit}
\bibinfo{author}{Pykhtin, M.}, \bibinfo{author}{Dev, A.}, \bibinfo{year}{2002}.
\newblock \bibinfo{title}{Credit risk in asset securitizations: Analytical
  model}.
\newblock \bibinfo{journal}{Risk} \bibinfo{volume}{15},
  \bibinfo{pages}{S16--S20}.
\bibitem[{Rodr{\'\i}guez-Moreno and Pe{\~n}a(2013)}]{rodriguez2013systemic}
\bibinfo{author}{Rodr{\'\i}guez-Moreno, M.}, \bibinfo{author}{Pe{\~n}a, J.I.},
  \bibinfo{year}{2013}.
\newblock \bibinfo{title}{Systemic risk measures: The simpler the better?}
\newblock \bibinfo{journal}{Journal of Banking \& Finance}
  \bibinfo{volume}{37}, \bibinfo{pages}{1817--1831}.
\bibitem[{Rosen and Saunders(2010)}]{rosen2010risk}
\bibinfo{author}{Rosen, D.}, \bibinfo{author}{Saunders, D.},
  \bibinfo{year}{2010}.
\newblock \bibinfo{title}{Risk factor contributions in portfolio credit risk
  models}.
\newblock \bibinfo{journal}{Journal of Banking \& Finance}
  \bibinfo{volume}{34}, \bibinfo{pages}{336--349}.
\bibitem[{Sch{\"o}nbucher(2001)}]{schonbucher2001factor}
\bibinfo{author}{Sch{\"o}nbucher, P.J.}, \bibinfo{year}{2001}.
\newblock \bibinfo{title}{Factor models: Portfolio credit risks when defaults
  are correlated}.
\newblock \bibinfo{journal}{The Journal of Risk Finance} \bibinfo{volume}{3},
  \bibinfo{pages}{45--56}.
\bibitem[{Schwerter(2011)}]{schwerter2011basel}
\bibinfo{author}{Schwerter, S.}, \bibinfo{year}{2011}.
\newblock \bibinfo{title}{Basel iii's ability to mitigate systemic risk}.
\newblock \bibinfo{journal}{Journal of financial regulation and compliance}
  \bibinfo{volume}{19}, \bibinfo{pages}{337--354}.
\bibitem[{Tarashev et~al.(2010)Tarashev, Borio and
  Tsatsaronis}]{tarashev2010attributing}
\bibinfo{author}{Tarashev, N.A.}, \bibinfo{author}{Borio, C.E.},
  \bibinfo{author}{Tsatsaronis, K.}, \bibinfo{year}{2010}.
\newblock \bibinfo{title}{Attributing systemic risk to individual institutions}
  .
\bibitem[{Uhde and Michalak(2010)}]{uhde2010securitization}
\bibinfo{author}{Uhde, A.}, \bibinfo{author}{Michalak, T.C.},
  \bibinfo{year}{2010}.
\newblock \bibinfo{title}{Securitization and systematic risk in european
  banking: Empirical evidence}.
\newblock \bibinfo{journal}{Journal of Banking \& Finance}
  \bibinfo{volume}{34}, \bibinfo{pages}{3061--3077}.
\bibitem[{Van~der Voort(2007)}]{van2007factor}
\bibinfo{author}{Van~der Voort, M.}, \bibinfo{year}{2007}.
\newblock \bibinfo{title}{Factor copulas: External defaults}.
\newblock \bibinfo{journal}{The Journal of Derivatives} \bibinfo{volume}{14},
  \bibinfo{pages}{94--102}.

\end{thebibliography}
\end{document}